\documentclass{aa}  

\usepackage{amsmath}
\usepackage{graphicx}
\usepackage{graphics}
\usepackage{txfonts}
\usepackage{natbib}{}
\usepackage{xcolor}
\usepackage{url}

\bibpunct{(}{)}{;}{a}{}{,}

\def\teff{\ifmmode T_{\rm eff} \else $T_{\mathrm{eff}}$\fi}

\def\ltsima{$\buildrel<\over\sim$}
\def\lsim{\lower.5ex\hbox{\ltsima}}

\newcommand{\hii}{H~{\sc ii}}
\newcommand{\ha}{\ifmmode {\rm H}\alpha \else H$\alpha$\fi}
\newcommand{\hb}{\ifmmode {\rm H}\beta \else H$\beta$\fi}
\newcommand{\hg}{\ifmmode {\rm H}\gamma \else H$\gamma$\fi}
\newcommand{\lya}{\ifmmode {\rm Ly}\alpha \else Ly$\alpha$\fi}

\newcommand{\Heiiuv}{He~{\sc ii} $\lambda$1640}
\newcommand{\Heiiopt}{He~{\sc ii} $\lambda$4686}

\newcommand{\ebv}{\ifmmode E_{\rm B-V} \else $E_{\rm B-V}$\fi}
\newcommand{\av}{\ifmmode A_{\rm V} \else $A_{\rm V}$\fi}

\def\kms{km s$^{-1}$}

\def\cmc{cm$^{-3}$}

\def\msun{\ifmmode M_{\odot} \else M$_{\odot}$\fi}
\def\msunyr{\ifmmode M_{\odot} {\rm yr}^{-1} \else M$_{\odot}$ yr$^{-1}$\fi}
\def\zsun{\ifmmode Z_{\odot} \else Z$_{\odot}$\fi}

\def\lsun{\ifmmode L_{\odot} \else L$_{\odot}$\fi}

\def\mup{\ifmmode M_{\rm up} \else M$_{\rm up}$\fi}
\def\mlow{\ifmmode M_{\rm low} \else M$_{\rm low}$\fi}

\def\aap{A\&A}

\def\apjl{ApJL}
\def\apjs{ApJS}

\newcommand{\oh}{\ifmmode 12 + \log({\rm O/H}) \else$12 + \log({\rm
O/H})$\fi}

\newcommand{\oii}{[O~{\sc ii}]}

\def\Oii{[O~{\sc ii}] $\lambda$3727}
\def\Oiii{[O~{\sc iii}] $\lambda\lambda$4959,5007}
\def\Oiiib{[O~{\sc iii}]$\lambda 5007$}
\def\Oiiit{[O~{\sc iii}]$\lambda 4363$}
\newcommand{\Neiii}{[Ne~{\sc iii}] $\lambda$3869}

\def\flyf{\ifmmode f_{\rm Lyf} \else $f_{\rm Lyf}$\fi}
\def\pz{\ifmmode P(z) \else $P(z)$\fi}
\def\ki2{\ifmmode \chi^2 \else $\chi^2$\fi}
\def\zphot{\ifmmode z_{\rm phot} \else $z_{\rm phot}$\fi}

\newcommand{\xphot}{\ifmmode x_\gamma \else $v_\gamma$\fi}
\newcommand{\xobs}{\ifmmode x_{\rm obs} \else $x_{\rm obs}$\fi}
\newcommand{\xcmf}{\ifmmode x_{\rm CMF} \else $x_{\rm CMF}$\fi}
\newcommand{\vexp}{\ifmmode V_{\rm exp} \else $V_{\rm exp}$\fi}
\newcommand{\vmax}{\ifmmode V_{\rm max} \else $V_{\rm max}$\fi}
\newcommand{\nh}{\ifmmode N_{\rm HI} \else $N_{\rm HI}$\fi}
\newcommand{\dv}{\ifmmode \Delta v({\rm em-abs}) \else $\Delta v({\rm em}-{\rm abs})$\fi}

\def\fesc{\ifmmode f_{\rm esc} \else $f_{\rm esc}$\fi}
\def\fescrel{\ifmmode f_{\rm esc,rel} \else $f_{\rm esc,rel}$\fi}

\def\frellya{\ifmmode f^{\rm rel}_{\rm{Ly}\alpha} \else $f^{\rm rel}_{\rm{Ly}\alpha}$\fi}

\def\hii{H{\sc ii}}

\newcommand{\mstar}{\ifmmode M_\star \else $M_\star$\fi}
\newcommand{\muv}{\ifmmode M_{1500} \else $M_{1500}$\fi}
\newcommand{\auv}{\ifmmode A_{\rm UV} \else $A_{\rm UV}$\fi}
\newcommand{\luv}{\ifmmode L_{\rm UV} \else $L_{\rm UV}$\fi}
\newcommand{\lir}{\ifmmode L_{\rm IR} \else $L_{\rm IR}$\fi}
\newcommand{\lbol}{\ifmmode L_{\rm bol} \else $L_{\rm bol}$\fi}
\newcommand{\liruv}{\ifmmode L_{\rm IR+UV} \else $L_{\rm IR+UV}$\fi}
\newcommand{\liroveruv}{\ifmmode L_{\rm IR}/L_{\rm UV} \else $L_{\rm IR}/L_{\rm UV}$\fi}
\newcommand{\nlyc}{\ifmmode N_{\rm Lyc} \else $N_{\rm Lyc} $\fi}
\newcommand{\rholyc}{\ifmmode \rho_{\rm Lyc} \else $\rho_{\rm Lyc} $\fi}
\newcommand{\chion}{\ifmmode \xi_{\rm ion} \else $\xi_{\rm ion}$\fi}
\newcommand{\chioncorr}{\ifmmode \xi_{\rm ion}^0 \else $\xi_{\rm ion}^0$\fi}

\newcommand{\Civuv}{C~{\sc iv} $\lambda$1550}

\newcommand{\Ciiiuv}{C~{\sc iii}] $\lambda$1909}
\newcommand{\Oiiiuv}{O~{\sc iii}] $\lambda$1666}
\newcommand{\Siiiiuv}{Si~{\sc iii}] $\lambda$1883,1892}
\newcommand{\Niiiuv}{N~{\sc iii}] $\lambda$1750}
\newcommand{\Nivuv}{N~{\sc iv}] $\lambda$1486}

\newcommand{\source}{CEERS-1019}

\begin{document}

\title{Extreme N-emitters at high-redshift: signatures of supermassive stars and globular cluster or black hole formation in action ?}
\subtitle{}
\author{R. Marques-Chaves\inst{1}, 
D. Schaerer\inst{1,2}, 
A. Kuruvanthodi\inst{1}, 
D. Korber\inst{1}, 
N. Prantzos\inst{3},
C. Charbonnel\inst{1,2}, 
A. Weibel\inst{1}, \\
Y. I. Izotov\inst{4},
M. Messa\inst{1,5}, 
G. Brammer\inst{6},
M. Dessauges-Zavadsky\inst{1}, 
P. Oesch\inst{1,6}
}
  \institute{Observatoire de Gen\`eve, Universit\'e de Gen\`eve, Chemin Pegasi 51, 1290 Versoix, Switzerland
\and CNRS, IRAP, 14 Avenue E. Belin, 31400 Toulouse, France
\and Institut d'Astrophysique de Paris, UMR 7095 CNRS, Sorbonne Université, 98bis, Bd Arago, 75014 Paris, France
\and Bogolyubov Institute for Theoretical Physics, National Academy of Sciences of Ukraine, 14-b Metrolohichna str., Kyiv, 03143, Ukraine  
\and The Oskar Klein Centre, Department of Astronomy, Stockholm University, AlbaNova, SE-10691 Stockholm, Sweden
\and Cosmic Dawn Center (DAWN), Niels Bohr Institute, University of Copenhagen, Jagtvej 128, K\o benhavn N, DK-2200, Denmark
       }

\authorrunning{Marques-Chaves et al.}
\titlerunning{Extreme N-emitters at high-redshift}

\date{Received date; accepted date}


\abstract{Recent JWST spectroscopic observations of the $z=10.6$ galaxy GN-z11 have revealed a very peculiar UV spectrum showing intense emission lines of nitrogen, which are generally not detected in galaxy spectra. This observation indicates a super-solar N/O abundance ratio at low metallicity, resembling only the abundances seen in globular cluster (GC) stars. This discovery suggests that we might be seeing proto-GCs in formation or possibly even signatures of supermassive stars.} 
{To examine if other objects with strong N{~\sc iv} and/or N{\sc iii} emission lines (N-emitters, hereafter) exist and to better understand their origin and nature, we have examined available JWST spectra and data from the literature.}
{Using the NIRSpec/JWST observations from CEERS we found an extreme N-emitter, \source\ at $z=8.6782$ showing intense \Nivuv\ and \Niiiuv\ emission.
From the observed rest-UV and optical lines we conclude that it is compatible with photoionization from stars and we determine accurate abundances for C, N, O, and Ne, relative to H.
We also (re-)analyze other N-emitters from the literature, including three lensed objects at $z=2.3-3.5$ (the Sunburst cluster, SMACS2031, and Lynx arc) and a low-redshift compact galaxy, Mrk 996. We compare the observed abundance ratios to observations from normal star-forming galaxies, predicted wind yields from massive stars and predictions from supermassive stars (SMS with $\sim 10^4-10^5$ \msun).}
{For \source\ we find a highly supersolar ratio $\log({\rm N/O})=-0.18 \pm 0.11$, and abundances of $\log({\rm C/O})= -0.75 \pm 0.11$ and $\log({\rm Ne/O})= -0.63 \pm 0.07$, which are normal compared to other galaxies at the low metallicity (\oh $= 7.70 \pm 0.18$) of this galaxy. The three lensed N-emitters also show strongly enhanced N/O ratios and two of them normal C/O. 
The high N/O abundances can be reproduced by massive star winds assuming a special timing and essentially no dilution with the ambient ISM. Alternatively, these N/O ratios can be explained by mixing the ejecta of SMS with comparable amounts of unenriched ISM. Massive star ejecta (from WR stars) are needed to explain the galaxies with enhanced C/O (Lynx arc, Mrk 996). On the other hand, SMS in the ``conveyer-belt model'' put forward to explain globular clusters, predict a high N/O and small changes in C/O, compatible with \source, the Sunburst cluster, SMACS2031, and GN-z11. 
Based on the chemical abundances, possible enrichment scenarios and other properties, such as their compactness and high ISM density, we discuss which objects could contain proto-GCs. We suggest that this is the case for \source, SMACS2031, and the Sunburst cluster. Enrichment in the Lynx arc and Mrk 996 is likely due to normal massive stars (WR), which implies that the star-forming regions in these objects cannot become GCs. Finally, we propose that some N-emitters enriched by SMS could also have formed intermediate mass black holes, and we suggest that this might be the case for GN-z11.
}
{Our observations and analysis reinforce the suggested link between some N-emitters and proto-GC formation, which is supported both by empirical evidence and quantitative models. Furthermore, the observations provide possible evidence for the presence of supermassive stars in the early Universe ($z>8$) and at $z \sim 2-3$. Our analysis also suggests that the origin and nature of the N-emitters is diverse, including also objects like GN-z11 which possibly host an AGN.}


 \maketitle

\section{Introduction}
\label{s_intro}

Known as the most distant spectroscopically-confirmed galaxy during several years \citep{Oesch2016},
GN-z11 has recently lead to new exciting and intriguing results, after the first spectra of this galaxy were obtained with the JWST. Indeed, the JWST/NIRSpec observations of \cite{Bunker2023JADES-NIRSpec-S} allowed to confirm a very high redshift of this source ($z=10.60$) and showed the presence of hydrogen, carbon, oxygen, magnesium, and neon emission lines in the rest-UV and rest-optical spectrum, often seen in star-forming galaxies at low-redshift and detected at $z \sim 4-8$ in other JWST spectra \citep[see e.g.,][]{Schaerer2022_SMACS,Cameron2023_JADES, Nakajima2023JWST-Census-for, Tang2023JWST/NIRSpec-Sp}.
Most surprisingly, however, the spectrum of GN-z11 revealed the presence of strong \Niiiuv\ and \Nivuv\ lines \citep{Bunker2023JADES-NIRSpec-S}, which are very rarely detected in galaxies \citep[see e.g.,][]{Barchiesi2022The-ALPINE-ALMA}. Furthermore, the object is found to be very compact \citep{Tacchella2023JADES-Imaging-o}, which could indicate the presence of massive compact star clusters or point to an active galactic nucleus (AGN) \citep{Bunker2023JADES-NIRSpec-S, Tacchella2023JADES-Imaging-o,Charbonnel2023N-enhancement-i,Maiolino2023_GNz11_AGN}.

The discovery of the peculiar emission line spectrum has triggered a series of papers discussing in particular their origin and the nature of GN-z11.
\cite{Bunker2023JADES-NIRSpec-S} first suggested that the strong N emission lines may imply an unusually high N/O abundance. They also discussed whether the emission would be powered by star formation or photoionization from an AGN, without reaching clear conclusions on this issue. The quantitative analysis of the emission line spectrum of GN-z11 by \cite{Cameron2023Nitrogen-enhanc} confirmed the high N/O abundance, with a lower limit of four times solar, finding also possibly a less extreme C/O ratio, and a metallicity (O/H), which is sub-solar, although not well constrained.
Using a suite of photoionization models, \cite{Senchyna2023GN-z11-in-conte} inferred the N/O abundance with a lower uncertainty and constrained the metallicity to $\oh = 7.84^{+0.06}_{-0.05}$, confirming in particular a large overabundance of N/O $\approx 3 \times$ solar.

The finding of an exceptionally high N/O abundance at low metallicity (typically ten times the normal N/O value at this O/H) has triggered different speculations about the sources and processes explaining this enrichment. The scenarii discussed include enrichment from massive stars winds (WR stars) or AGB stars, i.e.\ relatively ``classical scenarii'', or more ``exotic'' options such as pollution from PopIII star-formation, tidal disruption of stars from encounters with black holes, ejecta from very massive stars formed through collisions in dense clusters, and supermassive stars \citep[see:][]{Cameron2023Nitrogen-enhanc,Watanabe2023EMPRESS.-XIII.-,Senchyna2023GN-z11-in-conte,Charbonnel2023N-enhancement-i,Nagele2023Multiple-Channe}. Supermassive stars, for example, have been invoked by \cite{Charbonnel2023N-enhancement-i} and \cite{Nagele2023Multiple-Channe} since very strong enrichment of N and low metallicity is difficult to explain and requires fairly fined-tuned conditions with classical scenarios \citep[see also][]{Cameron2023Nitrogen-enhanc,Watanabe2023EMPRESS.-XIII.-}. Furthermore, such stars (with masses $\mstar \ga 1000$ \msun) have been proposed to form by runaway collisions in very dense stellar clusters, and they could explain the long-standing problem of multiple stellar populations and peculiar abundance patterns observed in globular clusters (GC), as discussed by \cite{Gieles2018Concurrent-form} and \cite{Denissenkov2014}. If correct, this would probably represent the first observational evidence of supermassive stars, which are also of great interest, for example for understanding the seeds of supermassive black holes \citep[e.g.,][and references therein]{2002ApJ...576..899P,2019PASA...36...27W,2023MNRAS.519.4753T}.

Abundance ratios are not the only properties observed in GN-z11 that resemble those of GCs. 
Its compactness and high ISM density also indicate conditions expected in young very massive clusters, which could be proto-GCs \citep{Senchyna2023GN-z11-in-conte,Charbonnel2023N-enhancement-i}.
GN-z11 might thus also be the first high-redshift object where the long sought-for peculiar abundance patterns characterizing GCs are observed 
\citep[e.g.,][and references therein]{2017MNRAS.469L..63R,2019A&ARv..27....8G}.
These exciting and surprising findings obviously pose the question of the uniqueness of GN-z11, beg for more examples, and call for a better understanding of similar objects, if they exist.

Indeed, although very rare, other galaxies showing emission lines of \Niiiuv\ or \Nivuv\ in the UV (referred to as N-emitters subsequently) are known, as pointed out by \cite{Senchyna2023GN-z11-in-conte} and found in the compilation of \cite{Barchiesi2022The-ALPINE-ALMA}. Apart from objects clearly identified as AGN, the Lynx arc, a lensed $z=3.36$ galaxy identified for \Nivuv\ and \Heiiuv\ emission is probably the first N-emitter studied in detail \citep{Fosb03,Villar-Martin2004Nebular-and-ste}. From photoionization modeling \cite{Villar-Martin2004Nebular-and-ste} derive a high N/O ratio and sub-solar metallicity.
Another strongly lensed object at $z=2.37$, the multiply-imaged compact star cluster in the Sunburst arc which has extensively been studied in recent years \citep[e.g.][]{Rivera-Thorsen2019Gravitational-l,Vanzella2022_sunburst}, shows \Niiiuv\ emission, as shown in the high S/N spectrum of \cite{Mestric2022_clumps}. \cite{Pascale2023Nitrogen-enrich} have shown that N/O is also elevated ($\sim 4 \times$ solar) at a metallicity $\sim 1/5$ solar.
Finally, in the low-redshift Universe, Mrk 996 uniquely stands out as the only galaxy showing strong \Niiiuv\ emission in the UV \citep[see][]{Mingozzi2022CLASSY-IV:-Expl}, and this blue compact dwarf galaxy has long been known as very peculiar, showing e.g.\ a high electron density, the presence of strong emission lines from WR stars in the optical, and a high N/O abundance, at least in its core \citep[e.g.][]{Thuan1996Hubble-Space-Te,James2009A-VLT-VIMOS-stu,Telles2014A-Gemini/GMOS-s}.

Here we present a detailed analysis of the $z=8.68$ galaxy \source\ observed with NIRSpec/JWST by the public CEERS survey \citep{Finkelstein2017jwst}. This object has previously been studied by several authors \citep{Tang2023JWST/NIRSpec-Sp,Nakajima2023JWST-Census-for,Larson2023A-CEERS-Discove}, but none of these have analysed the carbon and nitrogen abundance and its rest-UV spectrum.  
Only very recently, \cite{Isobe2023JWST-Identifica} have analyzed the UV spectrum in detail. Similarly to GN-z11, this galaxy exhibits a very peculiar rest-UV spectrum, making it clearly an N-emitter. Showing numerous emission lines of H, C, N, O, Ne, and the auroral \Oiiit\ line, it allows us to accurately determine the chemical abundances of these elements and offers thus a unique opportunity to study the second N-emitter in the early Universe and to enlarge the sample of these rare objects. 
We also analyze the other known N-emitters and compare their properties to those of \source\ and GN-z11. Finally, we confront the observed abundance patterns with predictions from normal massive stars and with predicted enrichment patterns from supermassive stars.

The paper is structured as follows. In Sect.\ \ref{s_obs} we describe the observational data, reduction, and measurements used in this work. We then discuss the nature of the ionizing source of \source\ (Sect.~\ref{s_agn}). The chemical abundances and other physical properties of \source\ are derived in Sect.~\ref{s_props}. In Sect.~\ref{s_discuss} we compare the abundance ratios of \source\ to other N-emitters and normal star-forming galaxies, and we present different chemical enrichment scenarios to explain them. We also discuss the possible link between \source\ and proto-GCs. The main results of our work are summarized in Sect.~\ref{s_conclude}. Throughout this work, we assume concordance cosmology with $\Omega_{\rm m} = 0.274$, $\Omega_{\rm \Lambda} = 0.726$, and $H_{0} = 70$ \kms\ Mpc$^{-1}$.
%

\section{CEERS-1019: a new strong N emitter at high redshift}
\label{s_obs}

CEERS-1019 ($\alpha$,~$\delta$ [J2000] = 215.0354$^{\circ}$, 52.8907$^{\circ}$) was initially identified as a $z_{\rm phot} \simeq 8.6$ dropout galaxy by \cite{Roberts-Borsani2016} and spectroscopically confirmed at $z_{\rm spec} = 8.683$ by \cite{Zitrin2015} through strong Ly$\alpha$ emission (see also \citealt{Mainali2018_NV1240} and \citealt{Witten2023_lya}). It is one of the most distant Ly$\alpha$ emitter known and is thought to reside in an over-dense region and ionized bubble boosting substantially its Ly$\alpha$ transmission \citep[][]{Larson2022_enviro, Leonova2022_enviro, Whitler2023_overdensityCEERS1019}. \cite{Mainali2018_NV1240} also report a tentative detection of N~{\sc v}~$\lambda 1240$ ($4.6\sigma$), suggesting a hard ionizing spectrum of this source.

Recently, much deeper spectroscopy of CEERS-1019 was reported and analyzed by \cite{Tang2023JWST/NIRSpec-Sp}, \cite{Nakajima2023JWST-Census-for}, and \cite{Larson2023A-CEERS-Discove} using NIRSpec, along with NIRCam and MIRI imaging. Although with some discrepancies, these works derived important physical properties of CEERS-1019 such as its stellar mass (log($M_{\star}/M_{\odot} \simeq 8.7-10.1$), gas-phase metallicities (\oh $ \simeq 7.6-8.0$), ionizing indicators (e.g., O32 $\simeq 13-18$), among others. Interestingly, \cite{Larson2023A-CEERS-Discove} reported a tentative ($2.5\sigma$) detection of a broad component in H$\beta$ that could be related to AGN activity (the presence of an AGN will be further discussed in Section \ref{s_agn}). Here, we re-analyze the available JWST data of CEERS-1019.

\subsection{JWST NIRSpec and NIRCam observations}

\begin{figure*}[htb]
\centering
\includegraphics[width=0.95\textwidth]{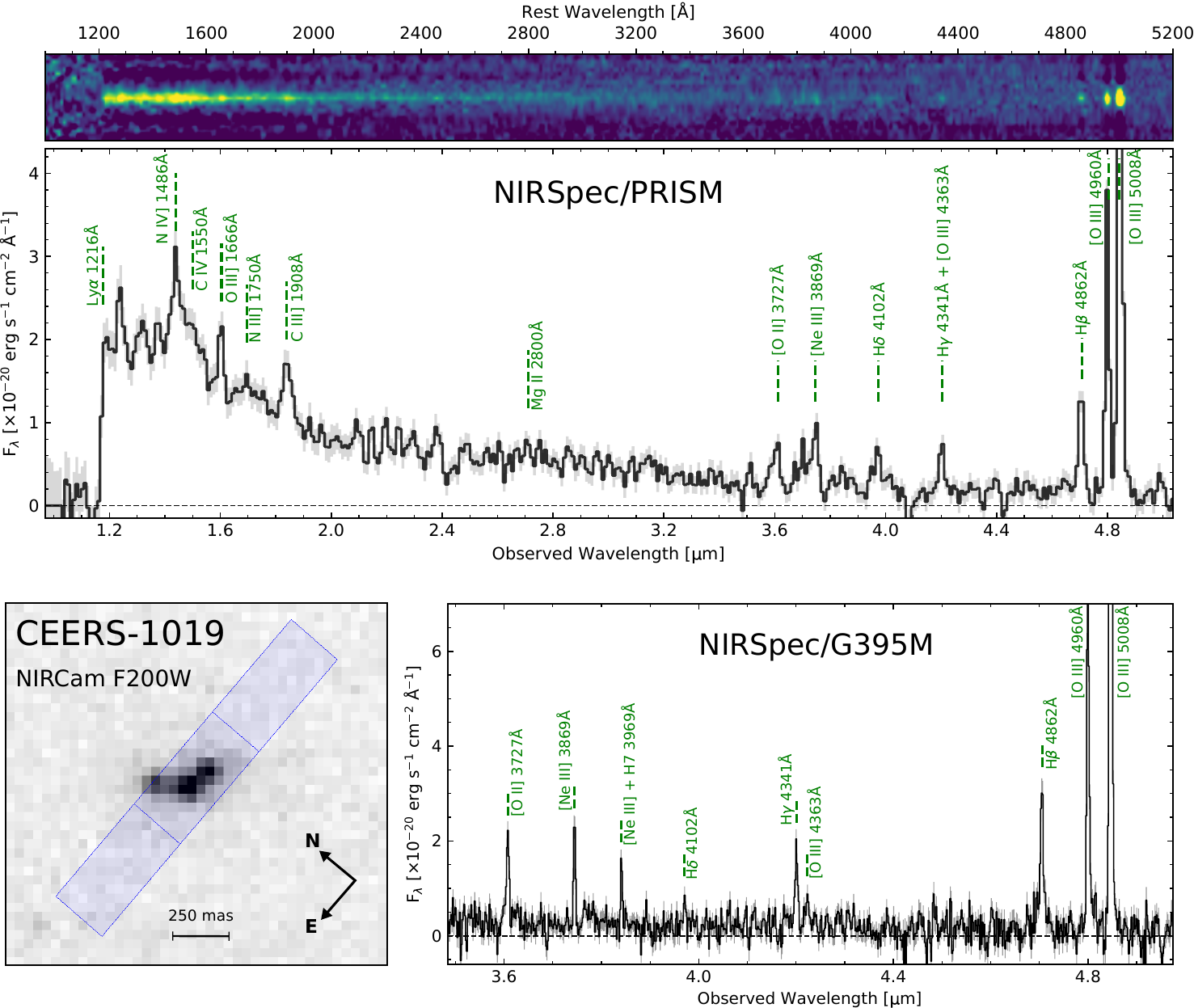}
\caption{Overview of the \textit{JWST} observations of \source{ }at $z=8.6782$. Top: 1D and 2D low-resolution NIRSpec/PRISM spectra (black) and 1$\sigma$ uncertainty (grey). Vertical dashed lines (green) mark the position of well-detected nebular emission lines. 
The X-axis in the bottom and top panels refer to the observed ($\mu$m) and rest-frame wavelengths (\AA), respectively. Bottom right: NIRSpec/G395M medium-resolution spectrum of \source. Bottom left: \textit{JWST} NIRCam cutout of \source{ }in the F200W filter. \source{ }is composed of three resolved clumps. The inferred positions of the NIRSpec MSA shutters are overlaid in blue.}
\label{fig_spec}
\end{figure*}

JWST/NIRSpec spectra are available for CEERS-1019 as part of the Cosmic Evolution Early Release Science (CEERS\footnote{\url{https://ceers.github.io/}}; \citealt{Finkelstein2022arXiv221105792F}) program. These observations include both low-resolution PRISM and medium-resolution grating (G140M/F100LP, G235M/F170LP, and G395M/F290LP), providing spectral resolution of $R\simeq 100$ and $R\simeq 1000$, respectively, and a spectral coverage $\simeq 1-5\mu$m. Standard 3-shutter slits and a 3-point nodding pattern were used. The total exposure time for each medium-resolution grating was 3107 seconds, split into three individual exposures of 14 groups each. Deeper observations were obtained with the low-resolution PRISM, with a total exposure time of 6214 seconds. Both PRISM and medium-resolution observations were obtained with an aperture position angle $\rm PA \simeq 89.32$ deg (see Figure \ref{fig_spec}).

Data reduction was performed using the official JWST pipeline \footnote{\url{https://jwst-pipeline.readthedocs.io/}} for Level 1 data products and {\sc msaexp}\footnote{\url{https://github.com/gbrammer/msaexp}} for Levels 2 and 3. Bias and dark current are subtracted followed by the correction of the $1/f$ noise and the “snowball” events. We use the calibration reference data system (CRDS) context {\sc jwst\_1063.pmap} to correct spectra for flat-field and implement the wavelength and photometric calibrations. 2D spectra of each slitlet are then drizzle-combined and the background is subtracted following the three-shutter dither pattern. 
Finally, 1D spectra are extracted using the inverse-variance weighted kernel following \cite{Horne1986}. Figure \ref{fig_spec} shows the NIRSpec spectra of CEERS-1019.

CEERS-1019 was also observed with \textit{JWST}/NIRCam with the F115W, F150W, F200W, F277W, F356W, F410M, and F444W filters with exposure times of $\sim 3000$ seconds \citep{Finkelstein2022arXiv221105792F}.  
NIRCam images were reduced using the {\sc grizli} reduction pipeline \citep{Brammer2023_GRIZLI_v1p82}, which includes procedures for masking the “snowball” artifacts and minimizing the impact of $1/f$ noise. Photometry of \source{ }is performed using SExtractor \citep{Bertin1996_SExtractor} in dual mode.
For each NIRCam filter, we use the point-spread functions (PSFs) provided by G. Brammer within the {\sc grizli} PSF library,\footnote{ \url{https://github.com/gbrammer/grizli-psf-library} } which are based on models from webbpsf \citep{Perrin2014_webbpsf}.
Images are then PSF-matched to F444W, which has the largest PSF within the NIRCam filters. We measure the flux of \source{ }in each filter using a circular aperture of $0.16^{\prime \prime}$ radius (4 pix) and apply an aperture correction derived in F444W using the "FLUX\_AUTO" measured in a Kron-like aperture with default Kron parameters. Then, we scale all fluxes to total fluxes based on the encircled energy of the circularized Kron aperture on the F444W PSF from webbpsf (see Weibel et al. in prep. for more details). As shown in the bottom left panel of Figure \ref{fig_spec}, \source{ }shows a complex morphology with three compact clumps.

\subsection{Emission line measurements}

As shown in Figure \ref{fig_spec}, \source\ presents intense nebular emission in the rest-frame UV and optical. As a first step, we determine the systemic redshift of \source\ using well-detected ($\geq 10\sigma)$ and uncontaminated (i.e., not blended) emission lines detected in the G395M spectrum. Using the centroids of \Neiii, \hg, \hb, and \Oiii\ we derive the mean value and scatter of $z_{\rm sys}=8.6782 \pm 0.0006$.

\begin{figure}[htb]
\centering
\includegraphics[width=0.48\textwidth]{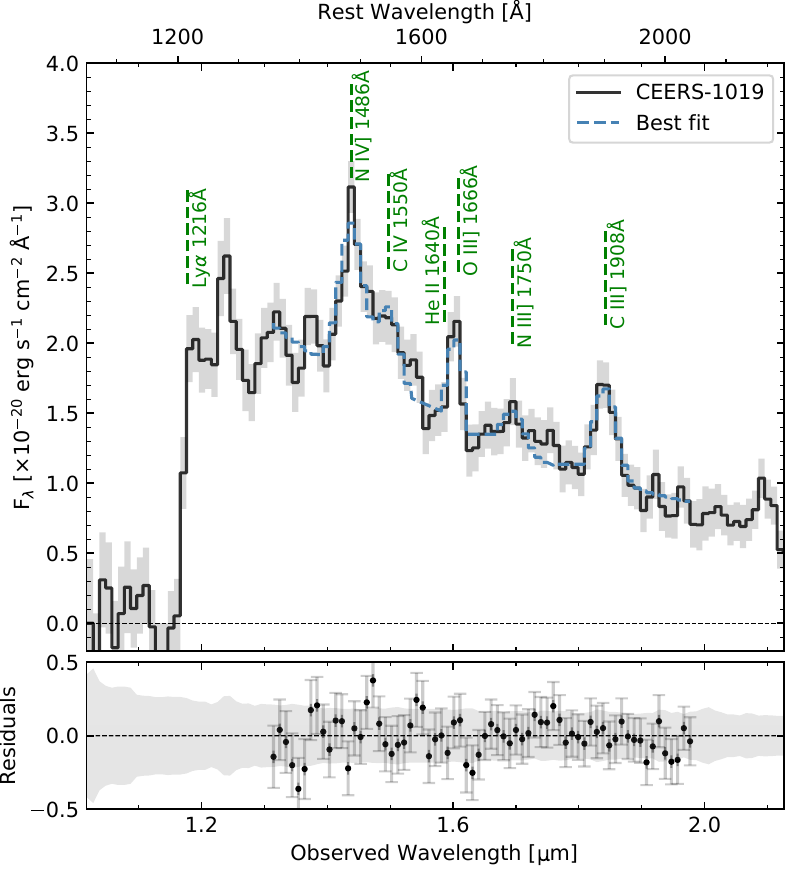}
\caption{Zoom-in to the rest-frame UV part of the PRISM spectrum of CEERS-1019 (black and 1$\sigma$ uncertainty in grey). Vertical dashed lines (green) mark the position of nebular emission lines. The blue line is the best fit for several UV emission lines and continuum. The bottom panel shows the residuals (in the same units as the spectrum) as well as the 1$\sigma$ uncertainties (grey).
}
\label{fig_spec_uv_prism}
\end{figure}

Several rest-frame UV lines are detected with high significance ($\geq 5\sigma$) in the deep PRISM spectrum (Figures \ref{fig_spec} and \ref{fig_spec_uv_prism}), such as \Nivuv\footnote{Unless stated otherwise, \Nivuv\ refers to the sum of the forbidden [N~{\sc iv}] $\lambda 1483$ and the semi-forbidden N~{\sc iv}] $\lambda 1486$ lines, which are not resolved in the Prism spectrum.}, \Civuv, \Oiiiuv, and \Ciiiuv. This contrasts with the shallower medium-resolution G140M spectrum that shows only Ly$\alpha$ and N~{\sc iv}] at $\geq 3\sigma$. Thus we use the much higher signal-to-noise ratio (S/N) PRISM spectrum to measure the fluxes of the rest-frame UV lines. 
We fit simultaneously several Gaussian profiles to account for the emission of N~{\sc iv}], C~{\sc iv}, O~{\sc iii}], N~{\sc iii}], and C~{\sc iii}], and a power-law in the form of $f_{\lambda} \propto \lambda^{\beta}$ to fit the continuum level between $1.3-2.1 \mu$m ($\lambda_{0} \simeq 1300-2200$\AA). Figure \ref{fig_spec_uv_prism} shows the results of the fit and the corresponding residuals. Since these lines are not resolved in the PRISM spectrum,\footnote{N~{\sc iv}] $\lambda 1486$ presents an observed line width $\rm FWHM = 394 \pm 95$ km s$^{-1}$ in the medium-resolution G140M spectrum.} we fixed the line widths of each line to the expected instrumental resolution at their corresponding wavelengths ($R\simeq 30-45$)\footnote{\url{https://jwst-docs.stsci.edu/jwst-near-infrared-spectrograph/nirspec-instrumentation/nirspec-dispersers-and-filters}}. We repeat the fit 500 times while bootstrapping the spectrum according to its $1\sigma$ error, and consider the standard deviation of each parameter as its $1\sigma$ uncertainty. Table \ref{tab_flux_measurements} summarizes our flux measurements. 
Along with Ly$\alpha$, N~{\sc iv}] is found to be the strongest emission line in the rest-UV, stronger than C~{\sc iv} and C~{\sc iii}] by a factor $\simeq 1.8$ and $\simeq 1.5$, respectively. We also infer a steep UV slope of $\beta_{\rm UV}^{\rm spec} = -2.11 \pm 0.09$ from the spectrum, which is consistent with the photometric one ($\beta_{\rm UV}^{\rm phot} = -2.11 \pm 0.15$) using the apparent magnitudes in the F150W and F200W filters ($\rm F150W=25.25\pm 0.08$ and $\rm F200W=25.29 \pm 0.07$).

\begin{table}
\begin{center}
\caption{Flux measurements of \source. \label{tab_flux_measurements}}
\begin{tabular}{l c c }
\hline \hline
\smallskip
\smallskip
Line & Flux & Grism/Prism \\
 & [$\times 10^{-18}$ erg s$^{-1}$ cm$^{-2}$] & \\
\hline 
Ly$\alpha$ $\lambda 1215$ & $2.38 \pm 0.49$ & G140M \\
N~{\sc v} $\lambda 1240$ & $<1.44$ (3$\sigma$) &  G140M \\
N~{\sc iv}] $\lambda\lambda  1483, 1486$ & $3.75\pm0.40$  & Prism \\
C~{\sc iv} $\lambda\lambda 1548,1550$ & $2.10\pm0.42$ & Prism \\
He~{\sc ii} $\lambda 1640$ & $<1.20$  (3$\sigma$)  & Prism \\
O~{\sc iii}] $\lambda\lambda 1661,1666$ & $1.64 \pm 0.32$  & Prism \\
\protect{N~{\sc iii}]} $\lambda 1750$ & $0.73\pm0.30$  & Prism \\
C~{\sc iii}] $\lambda\lambda 1907,1909$ & $2.43\pm0.36$ & Prism \\
\protect{[O~{\sc ii}]} $\lambda\lambda 3727,3729$ & $1.29\pm0.14$  & G395M \\
\protect{[Ne~{\sc iii}]} $\lambda 3869$ & $1.08\pm 0.16$  & G395M\\
H8+He~{\sc i} $\lambda 3889$ & $0.24\pm0.08$  & G395M\\
\protect{[Ne~{\sc iii}]} $+$ H7 $\lambda 3968$ & $0.60\pm0.11$ & G395M\\
H$\delta$ $\lambda 4101$ & $0.34\pm0.11$ & G395M\\
H$\gamma$ $\lambda 4340$ & $1.10\pm0.20$ & G395M\\
\protect{[O~{\sc iii}]} $\lambda 4363$ & $0.42\pm0.12$ & G395M\\
H$\beta$ $\lambda 4861$ & $2.14\pm0.22$ & G395M\\
\protect{[O~{\sc iii}]} $\lambda 4959$ & $4.50\pm0.24$ & G395M\\
\protect{[O~{\sc iii}]} $\lambda 5007$ & $14.05\pm0.28$ & G395M\\
\hline 
\end{tabular}
\end{center}
\end{table}

Flux measurements of rest-optical lines are obtained using the G395M spectrum, which presents a similar depth as the PRISM spectrum but with a much higher resolution. 
Optical lines are fitted separately over relatively narrow spectral windows (100 \AA, rest-frame) and a constant is assumed for the continuum level. The width of the lines is set as a free parameter. 
In total, we detect up to ten optical emission lines with high significance (Table \ref{tab_flux_measurements}), including Balmer lines that are useful for the determination of dust attenuation. 

To account for wavelength-dependent slit losses and absolute flux calibration, we derive the synthetic photometry of NIRSpec spectra (PRISM and gratings) through each NIRCam filter bandpass and matched it to that obtained from observed photometry. In this process, we use a wavelength-dependent polynomial function yielding scaling factors for the slit-loss correction ranging from approximately 2.0 (F150W) to 3.6 (F444W).

Using fluxes and equivalent widths of the detected Balmer lines \hb, \hg, and H$\delta$, we iteratively derive the dust attenuation $E(B-V) = 0.12\pm0.11$ using the \cite{Reddy2016} attenuation curve and following the methodology of \cite{Izotov1994}, which accounts for the internal extinction and underlying hydrogen stellar absorption. 
Other important lines, such as those that are sensitive to the electron temperature ($T_{\rm e}$, \Oiiit) and density ($n_{e}$, N~{\sc iv}] $\lambda\lambda 1483,1486$ and \oii$\lambda \lambda 3727,3729$) are also detected and are analyzed in more detail in Section \ref{s_props}. For the  N~{\sc iv}] and \oii{ }doublets we fit two Gaussian profiles with similar widths and use the expected separation between the two transitions. We find  line ratios of $F_{1483}/F_{1486} = 0.50\pm0.22$ and $F_{3727}/F_{3729} = 0.98\pm0.27$ for the N~{\sc iv}] and \oii{ }doublets, respectively.

We also check for the presence of spectral features that are usually associated with Wolf-Rayet (WR) stars. The so-called blue bump around 4600–4700\AA, encompassing the emission from N~{\sc iii} $\lambda 4640$, C~{\sc iii} $\lambda 4650$, and \Heiiopt, is detected neither in the G395M nor the PRISM spectra. We derive a $3\sigma$ upper limit relative to H$\beta$ of He~{\sc ii}/H$\beta \leq 0.26$. 
Similarly, the rest-UV \Heiiuv{ }line is not detected. Despite its low resolution, the PRISM spectrum clearly suggests no emission at the expected position of He~{\sc ii}, while the close O~{\sc iii}] emission is well detected (see Figure \ref{fig_spec_uv_prism}).

\section{The nature of the ionizing source: star formation versus AGN}\label{s_agn}

\begin{figure*}[htb]
\centering
\includegraphics[width=0.85\textwidth]{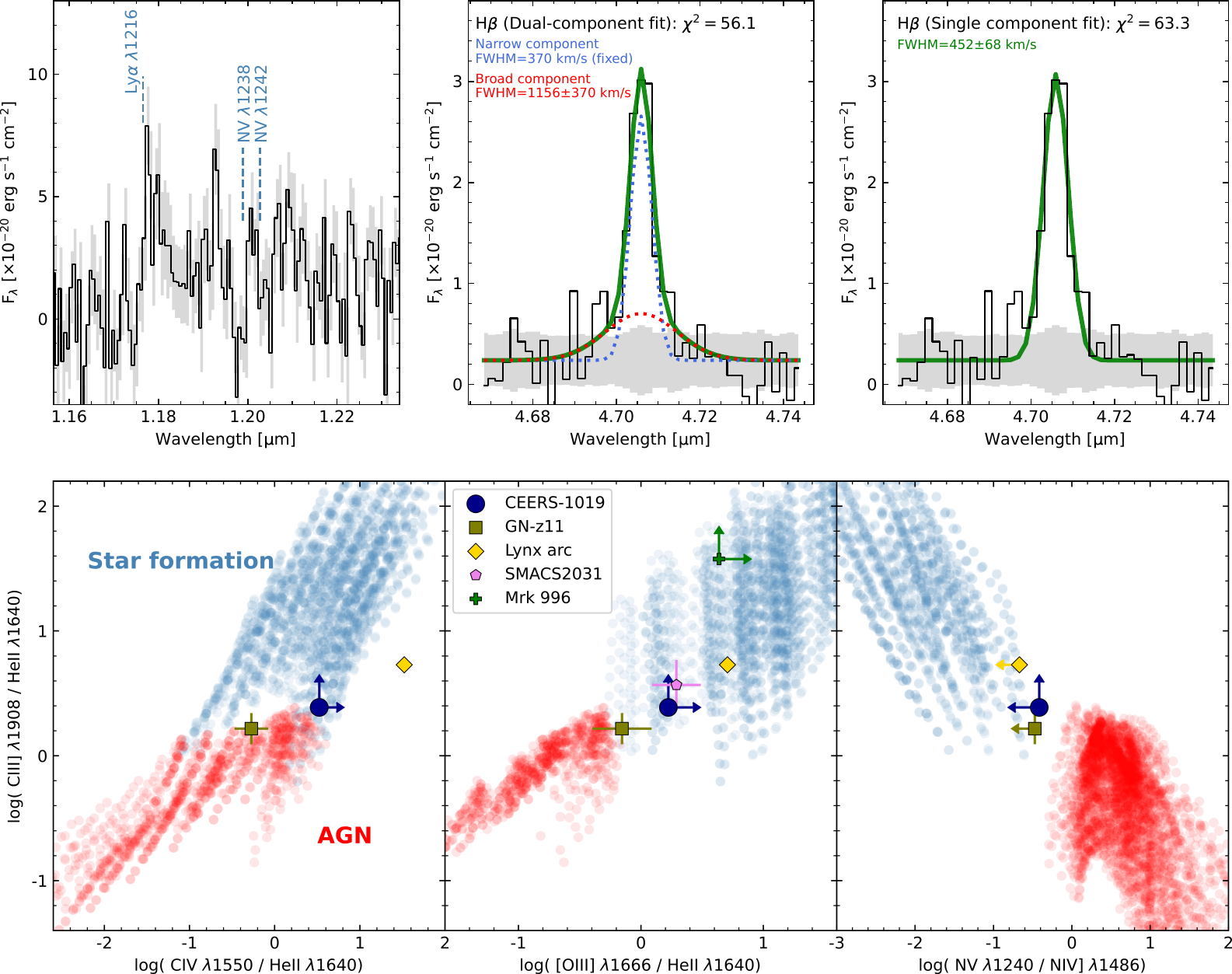}
\caption{Star formation and AGN diagnostics. The top left panel shows the G140M spectrum of \source{ }(in black and 1$\sigma$ uncertainty in grey) around the expected positions of N~{\sc v} $\lambda \lambda 1238,1242$ (marked with vertical lines), which are not detected ($1.44\times10^{-18}$ erg s$^{-1}$ cm$^{-2}$ at $3\sigma$). Ly$\alpha$ emission is also marked. The top middle and right panels show the best fits (green) to the H$\beta$ emission line using dual-component (middle, narrow and broad components in blue and red, respectively) and single-component (right) Gaussian profiles. The bottom panels show star formation (blue) and AGN (red) photoionization models using several rest-frame UV lines. The position of \source{ }(dark blue circle) aligns with the predictions of star-forming models in all diagnostic diagrams. The location of other known star-forming galaxies with strong Nitrogen emission, GN-z11 \citep{Bunker2023JADES-NIRSpec-S}, the Lynx arc \citep{Villar-Martin2004Nebular-and-ste}, SMACS2031 \citep{Patricio2016A-young-star-fo}, and Mrk~996 \citep{Mingozzi2022CLASSY-IV:-Expl} are also marked with different symbols as indicated in the legend. }
\label{fig_agn}
\end{figure*}

We now discuss the nature of the ionizing source of \source, building upon the recent findings by \cite{Mainali2018_NV1240} and \cite{Larson2023A-CEERS-Discove}, who suggest a possible AGN activity. In their study, \cite{Mainali2018_NV1240} reported the detection of N~{\sc v} $\lambda 1242$ emission with an integrated flux of $(2.8\pm0.6)\times 10^{-18}$~erg s$^{-1}$ cm$^{-2}$ with a narrow profile $\rm FWHM <90$~km s$^{-1}$ (unresolved in the MOSFIRE spectrum). However, the G140M spectrum does not exhibit any significant emission around the expected position of N~{\sc v} $\lambda \lambda 1238,1242$ (Figure \ref{fig_agn}, top left). 
By considering the flux uncertainty around $1.2\mu$m from the G140M error spectrum and assuming an unresolved line width of $\rm FWHM = 352$ km s$^{-1}$, we infer a $3\sigma$ limit of $1.44\times10^{-18}$ erg s$^{-1}$ cm$^{-2}$. This limit stands well below the reported value of \cite{Mainali2018_NV1240}. Furthermore, according to \cite{Morton1991_AtomicData}, N~{\sc v} $\lambda 1238$ is expected to be twice as strong as N~{\sc v} $\lambda 1242$ under standard conditions. Hence, considering the reported flux of \cite{Mainali2018_NV1240} for N~{\sc v} $\lambda 1242$, we would expect  $11.6\sigma$ and $5.8\sigma$ detections for N~{\sc v} $\lambda 1238$ and $\lambda 1242$, respectively. These limits, however, are incompatible with our observations.

\cite{Larson2023A-CEERS-Discove} reported a $2.5\sigma$ detection of a broad ($\simeq 1200$ km s$^{-1}$) component in H$\beta$ using the medium-resolution NIRSpec G395M spectrum. This broad component is not seen in stronger, forbidden lines like [O~{\sc iii}] $\lambda\lambda 4960,5008$, 
from which they suggest conditions similar to the broad line region (BLR) of an AGN. Using our own reduction of the G395M spectrum and a dual-component Gaussian profile to H$\beta$, we find a $2.2\sigma$ detection for the broad component (Figure \ref{fig_agn}, top middle). Clearly, deeper observations of H$\beta$ (or H$\alpha$ with MIRI) are needed to unambiguously confirm the presence and nature of the broad component in H$\beta$, as already discussed and suggested by \cite{Larson2023A-CEERS-Discove}. Indeed, if a single Gaussian profile is used to fit the H$\beta$ profile, a good fit is also found without penalizing significantly the residuals (Figure \ref{fig_agn}, top right). In this case, we find $\rm FWHM (H\beta) = 452\pm68$ km s$^{-1}$ which differs only by $1.2\sigma$ from the nominal $\rm FWHM = 369\pm16$ km s$^{-1}$ obtained for the much brighter [O~{\sc iii}] $\lambda 5008$ line.

If the existence of this broad component can be confirmed and attributed to the BRL, it would be expected that high-ionization semi-forbidden lines such as N~{\sc iv}], C~{\sc iv}, or C~{\sc iii}], which probe high-density regimes ($n_{\rm crit} \gtrsim 10^{9}$~cm$^{-3}$), would display similar broad Doppler widths as observed in type-1 AGNs \citep[e.g.,][]{Paris2011_QSO}.
However, these lines appear narrow in \source, especially N~{\sc iv}] which exhibits a high-significance detection and an intrinsic $\rm FWHM \simeq 160$ km s$^{-1}$ after correcting for instrumental broadening.
Thus, our results suggest that the aforementioned semi-forbidden lines are unlikely to originate from the broad line region. 
Instead, the properties of these lines, such as the narrow widths and the N~{\sc iv}] line ratio $F_{1483}/F_{1486} = 0.50\pm0.22$ (implying densities $n_e \approx 10^{4-5}$ \cmc, see Section \ref{density}), are consistent with narrow line regions of AGN or H~{\sc ii} regions. In the following, we discuss these two scenarios.

The lower panels of Figure \ref{fig_agn} present several diagnostic diagrams using different UV nebular lines: C~{\sc iii}]/He {\sc ii} versus C~{\sc iv}/He~{\sc ii}, [O~{\sc iii}]/He~{\sc ii}, and N {\sc v}/N {\sc iv}]. Photoionization models of star-forming galaxies from \cite{Gutkin2016} and narrow-line regions of AGN from  \cite{Feltre2016} are shown in blue and red, respectively. In the right panel of Figure \ref{fig_agn} we show models of star-forming galaxies from the updated BOND grid using Cloudy \citep{Ferland2017_cloudy}, which also includes N~{\sc iv}] and is available from the 3MdB\footnote{\url{https://sites.google.com/site/mexicanmillionmodels/}} \citep{Morisset2015A-virtual-obser}. These models encompass a wide range of parameters, including the ionizing parameter ($-4.0 \leq \log U \leq -1.0$), hydrogen number density ($10^{2} \leq n_{\rm H} \rm / cm^{3} \leq 10^{4}$), and the power law index of the ionizing spectrum ($-2.0 \leq \alpha \leq -1.2$). We have selected models with metallicities within the range $0.05 \leq Z/Z_{\odot} \leq 0.20$, which corresponds to the inferred metallicity for \source{} (12+log(O/H) $=7.70\pm0.18$, as indicated in Table \ref{ta_abund}). As illustrated in this figure, the position of \source\ (indicated by the blue circle) aligns with the predictions of star-forming models in all diagnostic diagrams. Clearly, the absence of He~{\sc ii} and N~{\sc v}, which probe energies $>54$ eV and $>77$ eV, respectively, places \source{ }far away from the region occupied by AGN models. It is worth noting that \cite{Isobe2023JWST-Identifica} suggested recently that the high N~{\sc iv}]/N~{\sc iii}] ratio observed in \source{ }is hardly reproduced by star formation models, pointing to an AGN contribution. However, the 3MdB photoionization models used here do predict very high ratios even well above the observed N~{\sc iv}]/N~{\sc iii}] $=5.1\pm2.2$, although requiring fairly high ionization parameters (log($U)\ga -2$). 

Other spectral features observed in \source, such as the intense N~{\sc iv}] emission compared to other UV lines (N~{\sc iv}]/C~{\sc iv} $\simeq 1.8$, N~{\sc iv}]/C~{\sc iii}] $\simeq 1.5$, N~{\sc iv}]/N~{\sc v} $\geq 2.6$), and narrow profiles ($\rm FWHM \simeq 160$ km s$^{-1}$ for N~{\sc iv}]) differ from those observed in AGNs, even those showing unusually strong Nitrogen lines \citep[e.g.,][]{Bentz2004_N_QSOs, Jiang2008_N_QSOs}. The so-called Nitrogen-loud QSOs exhibit much weaker N~{\sc iv}] compared to other lines (e.g., N~{\sc iv}]/C~{\sc iv}$\simeq 0.02-0.38$, \citealt{Batra2014_Nloud_QSOs}, \citealt{Dhanda2007Quasars-with-Su}) and, as expected, they present very broad Doppler widths ($\rm FWHM \simeq 1500-6000$ km s$^{-1}$, \citealt{Jiang2008_N_QSOs}). Similarly, some type-2 AGNs also present N~{\sc iv}] emission \citep[e.g.,][]{Hainline2011_AGN, Alexandroff2013_AGN},  but notably weaker compared to other high-ionization lines (N~{\sc iv}]/C~{\sc iv} $\simeq 0.15$, N~{\sc iv}]/C~{\sc iii}] $\simeq 0.34$, or N~{\sc iv}]/N~{\sc v} $\simeq 0.30$; \citealt{Hainline2011_AGN}). An exception may be GS-14, a type 1.8 AGN at $z\simeq 5.55$ recently analyzed by \cite{Ubler2023_Nemitter_AGN}. GS-14 exhibits broad components in Hydrogen and Helium lines ($\rm FWHM \simeq 3400$~km s$^{-1}$, \citealt{Ubler2023_Nemitter_AGN}) as well as narrow N~{\sc iv}] emission ($\rm FWHM \simeq 430$~km s$^{-1}$, \citealt{Vanzella2010_GS14}, \citealt{Barchiesi2022The-ALPINE-ALMA}), but it also shows clear nebular emission in N~{\sc v} $\lambda 1240$ and O~{\sc vi} $\lambda 1033$ \citep{Grazian2020_GS14, Barchiesi2022The-ALPINE-ALMA} which are not detected in \source.

In contrast, the spectrum of \source{ }resembles those of other, yet also rare star-forming galaxies with intense emission in Nitrogen lines. Examples such as the Lynx arc \citep{Fosb03, Villar-Martin2004Nebular-and-ste}, SMACS-2031 \citep{Christensen2012Gravitationally, Patricio2016A-young-star-fo}, Mrk 996 \citep{James2009A-VLT-VIMOS-stu, Mingozzi2022CLASSY-IV:-Expl}, and the Sunburst cluster show narrow and prominent N~{\sc iv}] and/or [N~{\sc iii}] lines suggestive of high electron temperatures and densities like \source{ }(see Section \ref{s_props}) and without any hint of AGN activity. The bottom panels of Figure \ref{fig_agn} also show the location of these strong N-emitters, all consistent with star-forming models like \source. The case of GN-z11, another strong N-emitter reported by  \cite{Bunker2023JADES-NIRSpec-S}, appears to be ambiguous, consistent with both models of AGN and star formation, as already discussed in \cite{Bunker2023JADES-NIRSpec-S} and \cite{Maiolino2023_GNz11_AGN}.
In conclusion, our results suggest that, regardless of the presence of an AGN whose confirmation awaits deeper data, the high-ionization lines observed in \source\ are consistent with stellar photoionization.

\section{Observational and derived physical properties of \source}
\label{s_props}

\subsection{ISM properties and element abundances}
The rich set of emission lines detected from the rest-frame UV-to-optical spectrum allows us to determine 
the electron temperature and density in the gas and the detailed abundances of numerous elements including 
H, C, N, O, and Ne. The derived quantities are summarized in Table \ref{ta_abund}.

\subsection{Electron temperature}
To derive physical conditions and element abundances we follow the prescriptions
of \cite{Izotov2006The-chemical-co}.
Briefly, these authors adopt the classical three-zone model of the
H~{\sc ii} region with electron temperatures $T_{\rm e}$(O~{\sc iii}) for the high-ionization zone,
and $T_{\rm e}$(O~{\sc ii}) for the low-ionization zone.  
The intermediate-ionization zone is not used here, since no such lines are detected.

The electron temperature $T_{\rm e}$(O~{\sc iii}) is derived both from the ratio of [O~{\sc iii}] line 
fluxes $\lambda$4363/$\lambda$(4959+5007) and from the  UV-to-optical line ratio of $\lambda$1660/$\lambda$5007.
The former ratio (rest-optical) is determined from the medium-resolution spectrum, the latter from the 
PRISM spectrum.
In both cases we obtain $T_e \approx 18000$ K, consistent within 1 $\sigma$, and with uncertainties between 1151 and 3252 K.
Subsequently, we adopt the electron temperature from the optical line ratios ($T_e=18849 \pm 3252$ K) with the larger uncertainty, which is primarily due to the low S/N detection of \Oiiit. 
The electron temperature in the low-ionization region is derived from relations obtained from the photoionization models of \cite{Izotov2006The-chemical-co}.

\subsection{Electron density}\label{density}

Several density indicators exist in the observed spectral range, but few can be used here in practice. 
In the UV, the \Ciiiuv, \Siiiiuv, and \Nivuv\ doublets are density estimators. 
However, the PRISM spectrum is of insufficient resolution to resolve any of these doublet lines.
\Siiiiuv\ is not detected, and \Ciiiuv\ has too low S/N in the medium-resolution spectrum.
Although of fairly low S/N, the \Nivuv\ doublet is detected with a ratio $\lambda$1483/$\lambda$1487 $=0.50 \pm 0.22$ which indicates a fairly high electron density of $n_e \approx 10^{4-5}$ \cmc\ \citep{Kewley2019_densities}.
In the optical, the \Oii\ doublet is clearly detected, but not resolved from the medium-resolution spectra. Our measured line ratio $\lambda$3727/$\lambda$3729 $= 0.98 \pm 0.23$ is consistent within the uncertainties with that obtained by \cite{Larson2023A-CEERS-Discove} ($0.639 \pm 0.255$), and compatible with $n_{e} > 10^{3}$ cm$^{-3}$ \citep{Kewley2019_densities}.

\begin{figure*}[htb]
\centering
\includegraphics[width=0.95\textwidth]{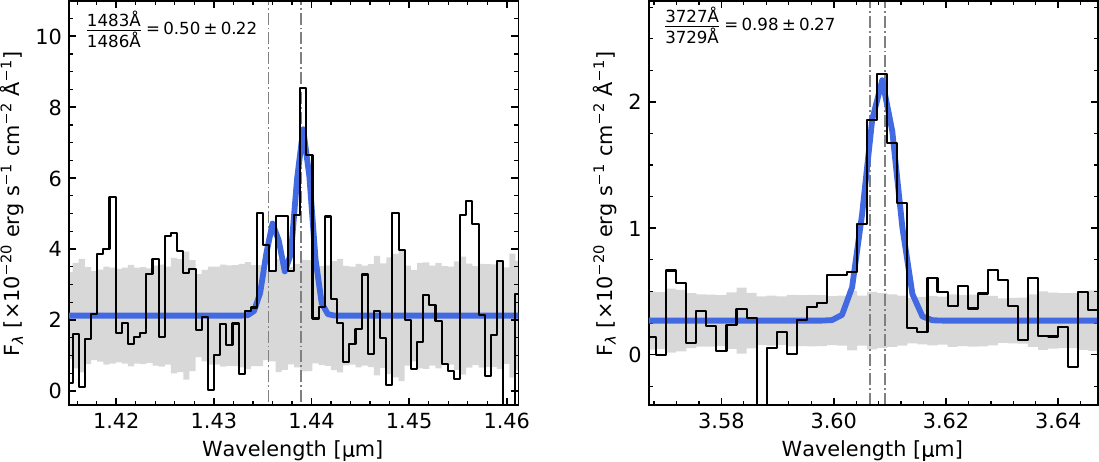}
\caption{Best fit (blue) of density-sensitive emission lines, N~{\sc iv}] $\lambda\lambda 1483,1486$ (left) and [O~{\sc ii}] $\lambda\lambda 3727,3729$ (right), using G140M and G395M medium-resolution spectra (black and $1\sigma$ uncertainty in grey), respectively. The fit uses two Gaussian profiles with similar widths and the expected position and separation between the two transitions (vertical lines).}
\label{fig_density}
\end{figure*}

The two density estimates could indicate a density gradient between the low and high ionization regions but are also compatible with a single, relatively high density of  $n_e \approx 10^{4-5}$ \cmc, whose origin we discuss below.
In any case, the most important point to take away from this is that the electron density, although high, is lower than the critical densities of all the relevant emission lines used for the subsequent abundance determinations.
This holds for the (semi-)forbidden lines of [O~{\sc iii}] at 1666, 4363, 4959, 5007 (with critical densities $n_{\rm crit} \ge 6.9 \times 10^5$ \cmc), the two components of the \Ciiiuv\ doublet ($n_{\rm crit} = 8.7 \times 10^4$ \cmc\ for 1907 and $10^9$ \cmc\ for 1909), \Civuv\ ($n_{\rm crit} = 2 \times 10^{15}$ \cmc), \Niiiuv\  (a multiplet whose components have $n_{\rm crit} \ge 10^9$ \cmc), \Nivuv\ ($n_{\rm crit} = 3 \times 10^9$ \cmc), and \Neiii\ ($n_{\rm crit} = 1 \times 10^8$ \cmc)
\citep[see e.g.][]{Hamann2002Metallicities-a,Dere2019CHIANTI}. 
Only the \Oii\ doublet, whose components have relatively low critical densities of $n_{\rm crit} = 1 (4) \times 10^3$ \cmc\ for 3728 (3726), is therefore affected by the high density inferred for \source, whereas all other lines can safely be used to determine abundances, to which we now proceed.

\begin{table}[ht]
\caption{ISM properties, ionic and total heavy element abundances }
\label{ta_abund}
\begin{center}
\begin{tabular}{lrrrrr}
\hline \hline
Property & Quantity \\ \hline 
$n_e$ [\cmc] & $10^4 - 10^5$ \\
$T_{\rm e}$(O {\sc iii}) -- UV [K] & $17839 \pm 1151$ \\
$T_{\rm e}$(O {\sc iii}) -- opt [K] & $18849 \pm 3252$\\
$T_{\rm e}$(O {\sc ii}) [K] &  $14937 \pm 944$\\
12+log(O$^{+}$/H$^+$) &  $6.73 \pm 0.14$ \\ 
12+log(O$^{2+}$/H$^+$) & $7.68 \pm 0.18$ \\
12+log(O/H)  &           $7.70 \pm 0.18$ \\
\\
log((N$^{2+}$+ N$^{3+}$)/O$^{2+}$)  -- UV only (V+04) & $-0.13 \pm 0.11$  \\
log((N$^{2+}$+ N$^{3+}$)/O$^{2+}$)  -- UV only (H+02) & $-0.16 \pm 0.17$  \\
log((N$^{2+}$+N$^{3+}$)/O$^{2+}$) -- UV+opt & $-0.18 \pm 0.28$  \\
\\
log((C$^{2+}$+ C$^{3+}$)/O$^{2+}$)  -- UV only (V+04) & $-0.75 \pm 0.11$  \\
log((C$^{2+}$+ C$^{3+}$)/O)  -- UV+opt (V+04) & $-0.79 \pm 0.22$  \\
log((C$^{2+}$+ C$^{3+}$)/O$^{2+}$)  -- UV only (PM17) & $-0.76 \pm 0.11$  \\
log(C$^{2+}$/O$^{2+}$)  -- UV only (I+23) & $-0.92 \pm 0.12$  \\
log(C$^{2+}$/O$^{2+}$)  -- UV+opt only (I+23) & $-0.84 \pm 0.22$  \\
ICF(C$^{2+}$/O$^{2+}$) = 1.1 \\
\\
ICF(Ne$^{2+}$/O$^{2+}$) & 1.04 \\
log(Ne/O)  &  $-0.63 \pm 0.07$ \\
\hline 
\end{tabular}
\end{center}
\end{table}%

\subsection{Ionic and total metal abundances}
\label{s_abund}
The electron temperature $T_{\rm e}$(O~{\sc iii}) is used to obtain abundances of ions O$^{2+}$, N$^{3+}$, N$^{2+}$, C$^{3+}$, C$^{2+}$,
and Ne$^{2+}$; the temperature in the low-ionization region, $T_{\rm e}$(O~{\sc ii}), to derive the ionic abundance of O$^+$.
Ionic abundances are derived following \cite{Izotov2006The-chemical-co} for the optical lines, and comparing different methods for the UV lines.
For C, N, and O, the observations provide two ionization stages, hence the ionic abundances will be
close to the total abundances, and we neglect further ionization corrections.
For Ne$^{2+}$ we use the ionization correction factor (ICF) following \citet{Izotov2006The-chemical-co}. The results are listed in Table \ref{ta_abund}.

We derive a total oxygen abundance of $\oh = 7.70 \pm 0.18$, which is dominated by the ionic abundance of O$^{2+}$/H$^+$ (see Table \ref{ta_abund}).
Given the high density, \Oii\ could be decreased and hence the O$^+$/H$^+$ abundance underestimated. 
However, in view of the high excitation observed from lines with high critical densities, it is likely that O$^{2+}$ is the dominant ionization stage over the majority of the \hii\ region and hence the determination of O/H close to the correct value.

With available line detections the N/O abundance can be determined in different ways. First we use only the UV lines to compute the ionic abundance ratio (N$^{2+}$+N$^{3+}$)/O$^{2+}$ using the expressions from \cite{Villar-Martin2004Nebular-and-ste} (V+04) and \cite{Hamann2002Metallicities-a} (H+02), assuming the low-density regime.  Then we determine N/H from the UV and optical line ratio (N and \hb) and use O/H determined from the optical lines. Both methods, marked as "UV only" and "UV+opt" respectively, yield values compatible within the errors, and consistent with a high N/O abundance $\log{(\rm N/O)} \approx -0.15 \pm 0.17$.

Similarly, for C/O we use the expressions from \cite{Villar-Martin2004Nebular-and-ste}, \cite{Perez-Montero2017Using-photo-ion} (PM17, and \cite{Izotov2023Abundances-of-C} (I+23) using either only the rest-UV or a combination of the UV and optical lines. As seen from Table \ref{ta_abund} the ionic abundance ratios derived in this manner are compatible within uncertainties. For the total C/O abundance we adopt $\log({\rm C/O})=-0.75\pm0.11$ as our default value. The C/O ratio is therefore clearly subsolar, and in fact very similar to the average of normal star-forming galaxies at the same O/H (see below). 
 
Finally, we also derive the Neon abundance from the \Neiii\ and \hb\ lines and applying an ICF from the oxygen lines, following \cite{Izotov2006The-chemical-co}.
We find an abundance ratio of $\log({\rm Ne/O}) = -0.63 \pm 0.07$, somewhat higher than the average value of $\log({\rm Ne/O}) = -0.78 \pm 0.01$ determined for normal star-forming galaxies by \cite{Guseva2011VLT-spectroscop} at the same metallicity.

Although the abundances derived here assume low densities they are not altered by density effects at the density derived for \source, as already discussed above. Most importantly, the critical densities for the \Niiiuv, \Nivuv, and \Oiiiuv\ lines involved in the (N$^{2+}$+N$^{3+}$)/O$^{2+}$ ratio derived from the UV are all very high ($n_{\rm crit} > 10^9$ \cmc), which further shows that this important ionic abundance ratio can be determined accurately.

Taken together, the derived abundances of \source\ show that this object has a ``metallicity'' (O/H) of approximately 1/10 solar \citep[asssuming a solar value of \oh=8.69][]{Asplund2009The-Chemical-Co,Asplund2021The-chemical-ma}, an exceptionally high N/O abundance, and a normal C/O abundance, when compared to galaxies of similar metallicity (see Fig.\ \ref{fig_abund}). The interpretation of these abundances and implications will be discussed below (Sect.\ \ref{s_discuss}).

\subsection{Comparison with other studies and caveats}
ISM properties and abundances of \source\ have been determined by several other studies, with whom we now briefly compared our results.

\cite{Larson2023A-CEERS-Discove} argue that the \Oii\ doublet can be deblended, from which they infer an electron density of $n_e = (1.9\pm0.2) \times 10^3$ \cmc. From inspection of the \Ciiiuv\ doublet, they suggest that the density could be higher than $n_e > 10^4$ \cmc. 
The density inferred here from the \Nivuv\ doublet ($n_e \approx 10^4 - 10^5$ \cmc) is compatible with their finding. Most importantly for the abundance determinations, all available density estimates indicate that the main emission lines should not be affected by density effects.

From their 3-$\sigma$ detection of \Oiiit\ \cite{Larson2023A-CEERS-Discove} inferred $T_e=18630 \pm 3682$ K, in excellent agreement with our determination. Based on the $T_e$ determination they infer $\oh=7.664 \pm 0.508$ from an average relation between $T_e$ and O/H determined empirically by \cite{Perez-Montero2017Using-photo-ion}.
\cite{Tang2023JWST/NIRSpec-Sp} determined $\oh=7.72^{+0.17}_{-0.14}$ using the direct method. Within the quoted uncertainties, our results agree with both of these determinations. A slightly higher O/H abundance ($\oh=7.97 \pm 0.16$), but still compatible with the uncertainties, has been derived by \cite{Nakajima2023JWST-Census-for} using a less accurate R23 strong-line calibration.
Finally, assuming AGN models, \cite{Isobe2023JWST-Identifica} have obtained a higher metallicity for \source, but similar N/O, C/O, and Ne/O ratios as derived here.

Note also that the abundance ratios determined here assume a homogeneous medium both in abundance and density. If pockets of high density and enriched gas coexist with lower density gas with say normal abundance ratios, only a relatively small fraction of enriched gas -- i.e.~relatively low amounts of Nitrogen -- might suffice to explain the observed emission line ratios since the emissivity of the forbidden line depends on the density \citep[see e.g][]{Izotov2006The-chemical-co}. However, in this case the inferred N/O abundance would also be a lower limit of the true N/O ratio in the enriched pocket.

\begin{figure*}[htb]
\centering
\includegraphics[width=1.0\textwidth]{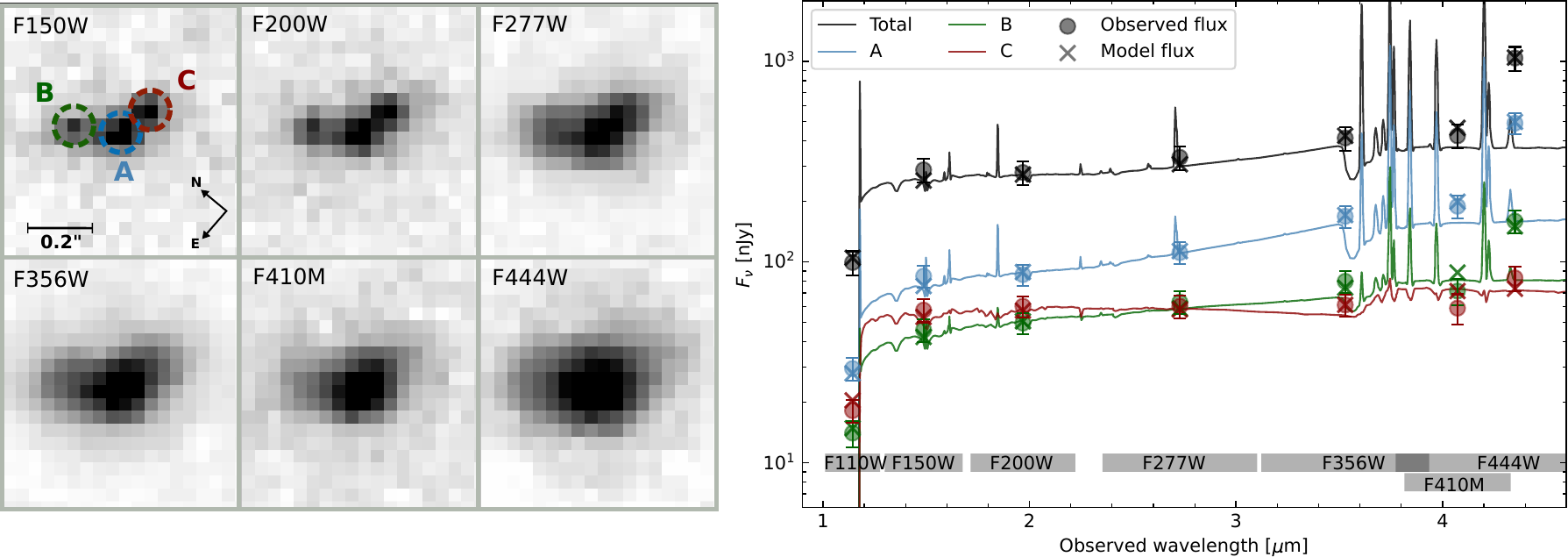}
\caption{Left panel shows cutout images around \source{ }in the NIRCam filters. In the F150W sub-panel, we show the positions of the three compact clumps resolved only at short wavelength, labeled as A, B, and C (blue, green, and red, respectively) The right panel shows the SED best-fit models using CIGALE \citep{Boquien2019_CIGALE} of the integrated light of \source{ }("Total" in black), as well as the individual clumps (A, B, and C in blue, green, and red, respectively). Observed fluxes are marked with circles, while the predicted fluxes from the best fit are marked with crosses. }
\label{fig_sed}
\end{figure*}

\subsection{Other physical properties}\label{sed}

\subsubsection{Morphology}

As shown in the left panel of Figure \ref{fig_sed} \source{ }shows a complex morphology in the NIRCam bands consistent with three different clumps/structures separated by $\simeq 0.24^{\prime \prime}$, or $\simeq 1.12$ kpc at $z=8.678$ ($4.68^{\prime \prime}$ kpc$^{-1}$). These clumps, labeled as A, B, and C as indicated in Figure \ref{fig_sed}, are very compact, only resolved in the NIRCam bands at short wavelengths. 

To investigate the morphology of \source{ }in more detail, we model the three galaxy substructures following accurately the methodology applied to the study of stellar clumps in \cite{Messa2022} and \cite{Claeyssens2023}. Assuming that clumps have Gaussian profiles, we consider a 15$\times$15 pixel region centered on the galaxy and we fit a model consisting of three 2D Gaussian functions, convolved to the NIRCam instrumental PSF in this field from the {\sc grizly} library. The best fit to their observed profiles (given by least-squares minimization) returns their fluxes and sizes. We assume that the shape of each substructure is the same in all bands. For this reason, the fit is initially performed in F200W, chosen as the reference filter, and then the shape (size, axis ratio, and position angle) of each clump is kept fixed in the other filters, where only the source flux is fitted. Uncertainties are obtained from Monte Carlo sampling.

The results of the model analysis are presented in Table \ref{tab_flux_components}. Our findings indicate that the morphologies of the three clumps in \source{ }are compact, with measured FWHMs of $48\pm5$ mas, $62\pm15$ mas, and $43\pm4$ mas for clumps A, B, and C, respectively. Following \cite{Peng2010_galfit} (see also:  \citealt{vanzella2017}, \citealt{Messa2022}, and \citealt{Claeyssens2023}), the inferred FWHM suggest that these clumps are resolved, albeit slightly, as their sizes are larger than the pixel size of the NIRCam images (40 mas). Translating these measurements into half-light radii, we find $r_{\rm e} = 112 \pm 12$ pc, $145\pm35$ pc, and $101\pm9$ pc for clumps A, B, and C, respectively.

\begin{table*}
\begin{center}
\caption{SED and morphological properties of the different substructures of \source. \label{tab_flux_components}}
\begin{tabular}{l c c c c c c c c}
\hline \hline
\smallskip
\smallskip
ID & R.A. & Dec. & $r_{\rm eff}$ & Age & SFR$_{10\rm Myr}$ & log($M_{\star}$) & log($\Sigma_{M}$) & log($\Sigma_{{\rm SFR}}$)\\
 & [J2000] & [J2000] & [pc] &  [Myr] & [$M_{\odot}$ yr$^{-1}$] & [$M_{\odot}$]  & [$M_{\odot}$ pc$^{-2}$] & [\msunyr kpc$^{-2}$]\\
(1) & (2) & (2) & (3) & (4) & (5) & (6) & (7) & (8)  \\
\hline 
A & 14:20:08.50 & $+$52:53:26.37 & $112\pm 12$ & $4.0\pm0.3$ & $148\pm25$ & $8.76\pm0.04$ & $3.86\pm0.11$ & $3.27\pm0.11$ \\

B & 14:20:08.51 & $+$52:53:26.51 & $145\pm35$ & $5.7 \pm 0.7$ & $83\pm18 $ & $8.66\pm0.15$ & $3.55\pm0.53$ & $2.81\pm0.21$ \\

C & 14:20:08.48  & $+$52:53:26.36 &  $101\pm9$ & $15.0\pm3.0$ & $ 2\pm10$ & $8.94\pm0.12$ & $4.14\pm0.14$ &$<2.27$ \\

Total & ---& ---  & --- & $14.4\pm7.2$ & $161\pm23$ & $9.31\pm0.15$ & --- & --- \\

\hline 
\end{tabular}
\end{center}
\textbf{Notes. ---} (1) different components of \source, (2) J2000 coordinates, (3) the effective radius, (4) burst age for components A, B, and C, and age assuming CSF for "total", (5) 10 Myr-weighted star-formation rate, (6) and (7) stellar mass and mass surface density, defined as $\Sigma_{M} = M_{\star} / (2 \pi r_{\rm eff}^{2})$, and (8) SFR surface density.

\end{table*}

\subsubsection{Spectral Energy Distribution}

We now analyze the spectral energy distributions (SEDs) of \source{ }as a whole (named Total) as well as its sub-components (A, B, and C). We use the SED-fitting code CIGALE \citep[][version 2022.1]{Boquien2019_CIGALE} using the available NIRCam photometry from F115W to F444W, covering the rest-frame wavelength $\sim 1200-4600$\AA. Stellar population models from \cite{bruzual2003} are used along with the \cite{chabrier2003} Initial Mass Function (IMF) and the Small Magellanic Cloud extinction curve ($R_{v}=2.93$, \citealt{Pei1992_SMC_law}). The metallicity is fixed to $Z=0.004$, the closest available value inferred for \source, and is assumed to be the same for nebular emission and starlight. The dust attenuation ($E(B-V)$) and ionization paramater (log($U$)) are treated as free parameters, ranging from $0.0-0.5$ mag and $-3.5$ to $-1.0$, respectively. Finally, we explore two different star-formation histories: a constant star-formation model applied to the integrated light of \source{ }(Total) and instantaneous burst episodes for the three sub-components (A, B, and C). For the former, we include the flux measurements of the H$\beta$ + [O~{\sc iii}] $\lambda\lambda 4960,5008$ emission lines in the fitting process.

Starting with the integrated emission of \source{ }(Total), the best-fit model, shown in black in the right panel of Figure \ref{fig_sed}, finds a continuous star-formation rate SFR$=161\pm23$ M$_{\odot}$ yr$^{-1}$ over $14\pm7$~Myr. The stellar mass is $M_{\star}^{\rm total}/ M_{\odot}=(2.0\pm0.6)\times 10^{9}$ attenuated by $E(B-V)=0.17\pm0.02$, in agreement with the values reported in \cite{Larson2023A-CEERS-Discove}.
For the three individual components A, B, and C we find burst masses of $M_{\star}^{\rm A}/ M_{\odot}=(5.7\pm0.5)\times 10^{8}$, $M_{\star}^{\rm B}/ M_{\odot}=(4.6\pm0.1)\times 10^{8}$, and $M_{\star}^{\rm C}/ M_{\odot}=(8.6\pm0.2)\times 10^{8}$, respectively. Clumps A and B are well-fitted with very young burst models, having ages of $4.0\pm0.26$ Myr and $5.6\pm0.7$ Myr, respectively. On the other hand, clump C is older than the other components, with a burst age of $15.0\pm 2.9$ Myr. Indeed, the color obtained for clump C $\rm F356W - F444W = 0.32 \pm 0.29$ is significantly lower than those measured in clumps A and B, $\rm F356W - F444W \simeq 0.75-1.16$, suggesting a weak contribution of nebular emission in F444W (e.g., H$\beta$ and [O~{\sc iii}]), thus negligible star formation over the last $\lesssim 10$~Myr. 

We note here, in passing, that our estimates of stellar mass and SFR are based on standard models, which do, e.g., not account for peculiar abundance patterns, and which rely on standard assumptions regarding the star-formation histories. This approach allows meaningful comparisons of these parameters with those from other studies. Also, for simplicity, we do not correct the photometry for a possible contribution from SMSs, since we expect that such an object does not dominate the light in \source. Indeed, even for a very massive SMS with $10^6$ \msun\ the expected flux in the rest-optical range is $m_{\rm AB} \sim 28$ \citep{Martins2020Spectral-proper}, approximately 10 times fainter than the flux of region A.

\subsubsection{Stellar mass and SFR surface densities}

Based on the stellar masses and half-light radii obtained for the individual clumps (Table \ref{tab_flux_components}), we obtained high stellar mass surface densities of log($\Sigma_{M})=3.86\pm0.11$, $3.55\pm0.53$, and $4.14\pm0.14$ ${\rm M_{\odot} pc^{-2}}$ for clumps A, B, and C, respectively (defined as $\Sigma_{M} = M_{\star} / (2 \pi r_{\rm eff}^{2})$). It is worth noting that the inferred values of $\Sigma_{M}$ may even be higher if each substructure comprises multiple unresolved stellar systems. Nevertheless, these values are already comparable to the densest systems identified at high redshift by \cite{Claeyssens2023} or \cite{Mestric2022_clumps}, and significantly higher than the average log($\Sigma_{M}) \simeq 2$ ${\rm M_{\odot} pc^{-2}}$ observed in nearby young clusters \citep{Brown2021_YSC}. Similarly, the compactness index, defined as {\bf $C_{5} = (M_{\star}/10^{5} M_{\odot}) / (r_{\rm eff}$/pc$^{-1})$} is also high in the case of \source. It ranges from $C_{5} \simeq 30-90$ depending on the clump, exceeding the values of old globular clusters and young massive clusters by at least one order of magnitude \citep{Krause2016_C5}, suggesting high cluster formation efficiencies \citep{Krause2016_C5, Kruijssen2012_Gamma}. 
The SFR surface density is also found to be very high for clumps A and B with log($\Sigma_{\rm SFR})=3.27\pm0.11$  and $2.81\pm0.21$ \msunyr kpc$^{-2}$, respectively. In contrast, clump C does not show significant star formation over the last 10~Myr, yielding an upper limit of log($\Sigma_{\rm SFR})<2.27$ \msunyr kpc$^{-2}$.

Finally, the derived mass and SFR surface densities in \source{ }are comparable with those of other prominent N-emitters discussed below, such as GN-z11 (log($\Sigma_{M}) \sim 4.6$ ${\rm M_{\odot} pc^{-2}}$, \citealt{Tacchella2023JADES-Imaging-o}), SMACSJ2031 (log($\Sigma_{M}) \sim 4.0$ ${\rm M_{\odot} pc^{-2}}$, log($\Sigma_{\rm SFR}) \sim 1.4$ \msunyr kpc$^{-2}$, \citealt{Patricio2016A-young-star-fo}), the Sunburst cluster (log($\Sigma_{M}) \sim 4.1$ ${\rm M_{\odot} pc^{-2}}$, log($\Sigma_{\rm SFR}) \sim 3.7$ \msunyr kpc$^{-2}$, \citealt{Vanzella2022_sunburst}), or Mrk 996 (log($\Sigma_{M}) \sim 2.8$ ${\rm M_{\odot} pc^{-2}}$, \citealt{Thuan1996Hubble-Space-Te}). This suggests a potential connection between compactness and a high production efficiency of nitrogen.

\subsection{Mass of the enriched material}
\label{subsec:mass_gas_obs}

The total mass of enriched, ionized gas, which is directly observable, can easily be estimated assuming ionization equilibrium and a constant ISM density \citep[see, e.g.,][]{Dopita2003}:
\begin{equation}
        M_{\rm ionized} = \frac{m_p Q_H}{\alpha_B n_e} = 2.5 \times 10^6 \left(\frac{10^3}{n_e}\right) \left(\frac{Q_H}{10^{54}}\right) \msun, 
\end{equation}
where $Q_H$ is the ionizing photon production rate which can be determined from H recombination lines, $n_e$ the electron density, $m_p$ the proton mass, and $\alpha_B$ the recombination rate coefficient. 

For \source\ we thus find $M_{\rm ionized} \sim 1.2 \times 10^5$ \msun, from the observed \hb\ luminosity and adopting $n_e=10^5$ \cmc, very similar to $M_{\rm ionized} \sim 2 \times 10^5$ \msun\ inferred for GN-z11 by \cite{Charbonnel2023N-enhancement-i}. \cite{Maiolino2023_GNz11_AGN} argue that the amount of enriched gas in GN-z11 could be even smaller if the N-emitting gas is found at higher densities, as they suggest.

\begin{figure}[tb]
\centering
\includegraphics[width=0.48\textwidth]{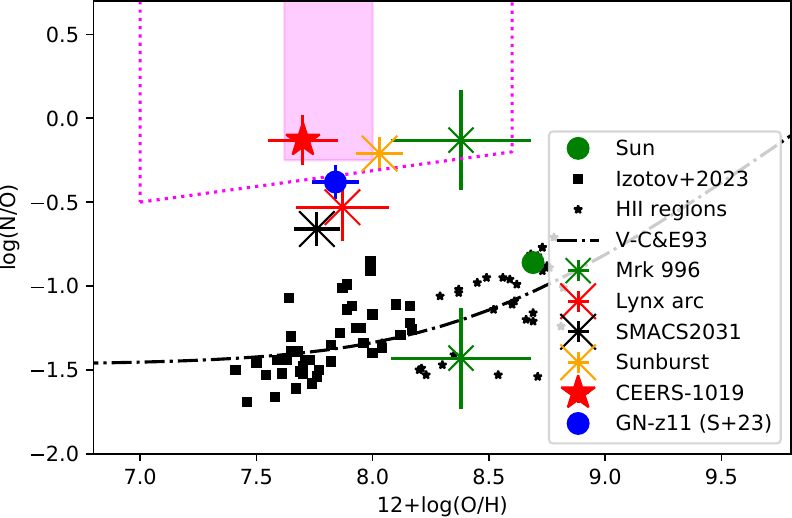}\vspace{0.5cm}
\includegraphics[width=0.48\textwidth]{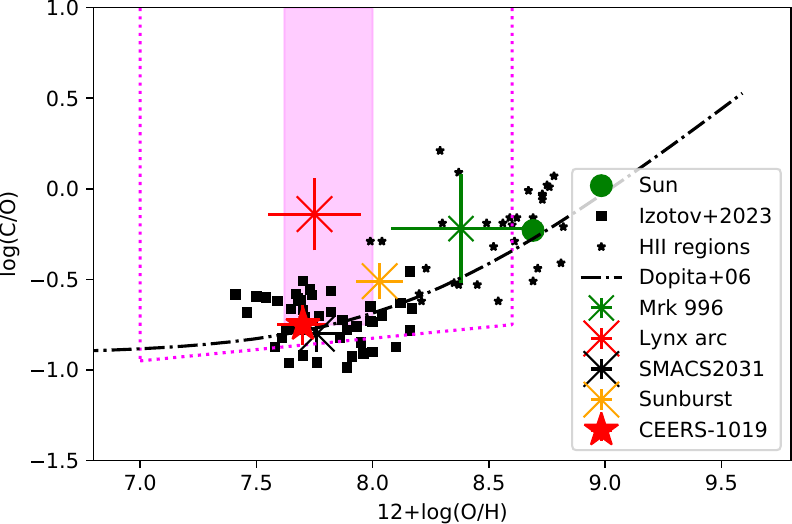}
\caption{Observed chemical abundances of the six N-emitters and comparison samples from the literature. {\em Top:} N/O versus O/H, {\em bottom:} C/O versus O/H.
\source\ is shown by a red star, GN-z11 by a blue circle. The $z \sim 2.6-3$ lensed galaxies (Lynx arc, SMACS2023 and the Sunburst cluster) are shown by red, black, and orange crosses, the low-$z$ galaxy Mrk 996 with a green cross (two N/O values from \cite{James2009A-VLT-VIMOS-stu} are shown: for the central region and from the total galaxy).
The magenta shaded region and outlined box indicate the range of abundances allowed for GN-z11, according to \cite{Cameron2023Nitrogen-enhanc}.
Low-$z$ star-forming galaxies and \hii\ regions from the compilation of \cite{Izotov2023Abundances-of-C} are shown by small black symbols.
The dash-dotted line shows the average trend observed in low-$z$ star-forming galaxies, as parametrized by \cite{Vila-Costas1993The-nitrogen-to} for N/O and C/O by \cite{Dopita2006Modeling-the-Pa} respectively.}
\label{fig_abund}
\end{figure}

\section{Discussion}
\label{s_discuss}

\subsection{Observed heavy element abundances in \source\ comparison to ``normal'' objects}

The main elemental abundance ratios derived for \source\ are shown in Fig.\ \ref{fig_abund}, and compared to measurements in other galaxies and \hii\ regions. To do so we use in particular the recent CNO abundances determined and compiled by \cite{Izotov2023Abundances-of-C}, who primarily included data from low-redshift star-forming galaxies observed with HST/COS, and data on individual \hii\ regions from the works of \cite{Esteban2002, Esteban2009, Esteban2014}, \cite{GarciaRojas2007}, and \cite{LopezSanchez2007}.

As well known, the majority of galaxies and \hii\ regions follow a fairly well-defined sequence of N/O versus O/H and C/O versus O/H \citep[e.g.][]{Garnett1999,Berg2019The-Chemical-Ev}, which can be understood with chemical evolution models \citep{Henry2000On-the-Cosmic-O,Chiappini2006,Prantzos2018MNRAS}. In N/O, for example, only few strong outliers with a large nitrogen excess are known at low redshift \citep[see e.g.][]{Thuan1996Hubble-Space-Te,Pustilnik2004HS-08374717---a,Stephenson2023}.
In comparison, \source\ clearly stands out by having an extremely high Nitrogen abundance, $\log({\rm N/O}) = -0.13 \pm 0.11$, which is approximately 5.6 times the solar ratio \citep{Asplund2021The-chemical-ma} and more than a factor 10 higher than the N/O values generally observed at similar metallicities (O/H). This exceptionally high N abundance reflects the very peculiar UV spectrum of \source, showing unusually strong Nitrogen lines. 

In contrast to N/O, with $\log({\rm C/O})=-0.75 \pm 0.11$, the C/O abundance is fairly normal for the observed metallicity. The Ne/O abundance, $\log({\rm Ne/O})=-0.63 \pm 0.07$ is somewhat higher (by $\sim 0.15$ dex) than the average value for normal star-forming galaxies derived by \cite{Guseva2011VLT-spectroscop}  at the same metallicity. 

Interestingly, these observed abundance ratios of \source\ resemble those of globular cluster stars, similarly to what was pointed out by \cite{Senchyna2023GN-z11-in-conte} and \cite{Charbonnel2023N-enhancement-i} for GN-z11. The origin of these peculiar abundances ratios will be discussed below.

\subsection{Abundances in other N-emitters}

Interestingly, the abundance ratios found in \source\ resemble those found by 
\cite{Cameron2023Nitrogen-enhanc} for the $z=10.6$ galaxy GN-z11 observed recently with JWST by \cite{Bunker2023JADES-NIRSpec-S}, which are shown by boxes in Fig.\ \ref{fig_abund}. 
As shown, the abundances in GN-z11 suffer from large uncertainties, which are in particular due to the fact that the \Oiiib\ line is shifted beyond the range accessible with NIRSpec and no direct O/H abundance determination is possible for this object from the present data.
Using photoionization modeling, \cite{Senchyna2023GN-z11-in-conte} have further constrained the abundances in GN-z11, obtaining total gas abundances of $\oh=7.84\pm0.06$ and
$\log({\rm N/O})=-0.38\pm0.05$, which are quite similar to those obtained here for \source.
Clearly, both \source\ and GN-z11 are significantly enriched in Nitrogen, reaching exceptionally high N/O values.
The carbon abundance cannot be well constrained in GN-z11, since the electron temperature remains undetermined in this object. The allowed range, derived by \cite{Cameron2023Nitrogen-enhanc}, is indicated in Fig.\ \ref{fig_abund}.

Very few other galaxies or \hii\ regions with a high N/O abundance and/or clear detections of nebular lines of N in the UV can be found in the literature.
\cite{Barchiesi2022The-ALPINE-ALMA} list known AGN and galaxies with O~{\sc vi}, N~{\sc v}, or \Nivuv\ emission lines in the rest-UV. Among the non-AGN in their list one finds the peculiar galaxy named the Lynx arc (at $z=3.36$), which has been studied by \cite{Fosb03} and \cite{Villar-Martin2004Nebular-and-ste}, although \cite{Binette2003High-z-nebulae} have argued that this object may be an obscured QSO. According to the photoionization models of \cite{Villar-Martin2004Nebular-and-ste}, both the N/O and C/O abundance ratios of this object are elevated, as seen in Fig.\ \ref{fig_abund}. Although suspected, no direct signs of WR stars have been found in this object \citep{Villar-Martin2004Nebular-and-ste} and the inferred abundances are not explained.

Another object showing nebular \Nivuv\ emission is the strongly lensed galaxy SMACSJ2031.8-4036 at $z=3.5$ studied in detail by \cite{Christensen2012Gravitationally} and \cite{Patricio2016A-young-star-fo}. The available VLT observations (with XShooter and MUSE) cover both the rest-UV and optical domain, allowing the detection of numerous emission lines, and thus electron temperature, density and abundance determinations. 
Interestingly, this object shows indications for high density ($n_e \ga 10^5$ \cmc) from the \Nivuv\ doublet and lower densities from other diagnostics \cite{Patricio2016A-young-star-fo}. The metallicity $\oh = 7.76 \pm 0.03$ is very similar to \source\ and it shows a normal C/O abundance ($\log({\rm C/O})=-0.80 \pm 0.09$), according to \cite{Christensen2012Gravitationally}. 
Inspection of their spectra, kindly provided by the authors, shows a clear detection of both \Nivuv\ and \Niiiuv\ lines, which allows us to determine N/O from the 
UV lines and the reported $T_e$ using the same methods described above (see Sect.\ \ref{s_abund}). We find a relatively high N abundance of $\log({\rm N/O})=-0.66 \pm 0.1$, which we also report in  Fig.\ \ref{fig_abund}.
Finally, we also find a normal Neon abundance of $\log({\rm Ne/O})=-0.82$ from the reported line fluxes.

In the list of \cite{Barchiesi2022The-ALPINE-ALMA} other non-AGN spectra showing UV lines of Nitrogen show only N~{\sc v} P-Cygni lines, which are most likely due to stellar emission, or are stacked spectra with weak detections, not suitable for our purpose. 

Another high-redshift object where \Niiiuv\ emission has recently been detected is the strongly lensed and multiply imaged stellar cluster at $z=2.368$ in the Sunburst arc \citep{Mestric2023Clues-on-the-pr}, an exceptional object studied in depth by various authors \citep[e.g.][]{Rivera-Thorsen2019Gravitational-l,Vanzella2020Ionizing-the-in}. From a detailed analysis and photoionization modelling, \cite{Pascale2023Nitrogen-enrich} infer in particular a high N/O abundance ratio ($\log {\rm N/O} = -0.21^{+0.10}_{-0.11}$), and normal C/O and Ne/O ratios for a metallicity (O/H) of approximately $\sim 0.22$ solar.
The N/O ratio of this object fares thus among the highest values, comparable to \source, and C/O is also similar, as also shown in Fig.~\ref{fig_abund}.

To extend our comparison, we have also examined the low-redshift galaxy Mrk 996, which is a well-known Blue Compact Dwarf (BCD) galaxy with peculiar properties, such as a high electron density, broad emission line components in \ha, \Oiii\ and other lines, the presence of Wolf-Rayet stars of WN and WC type, and a high N/O abundance \citep[see e.g.][]{Thuan1996Hubble-Space-Te,Pustilnik2004HS-08374717---a,James2009A-VLT-VIMOS-stu,Telles2014A-Gemini/GMOS-s}.
This galaxy also shows N~{\sc iii}] and N~{\sc iv}] emission lines in the UV \citep{Mingozzi2022CLASSY-IV:-Expl,Senchyna2023GN-z11-in-conte}.
From integral-field observations \cite{James2009A-VLT-VIMOS-stu} have found a normal N abundance ($\log({\rm N/O}) \approx -1.43$) across the galaxy and a N-enhancement by a factor $\sim 20$  ($\log({\rm N/O}) \approx -0.13$) in the broad line component, emitted in the central region. The two measurements are plotted in Fig.\ \ref{fig_abund}.
The C/O abundance of Mrk 996 can be derived from the \Ciiiuv\ and \Oiiiuv\ line ratio, which is taken from the 
HST/COS observations from the CLASSY survey \citep{Berg2022The-COS-Legacy-,Mingozzi2022CLASSY-IV:-Expl}, and adopting the electron temperature $T_e=10^4$ K from \cite{James2009A-VLT-VIMOS-stu}. We find a high Carbon abundance of $\log({\rm C/O})= -0.22$, close to solar, for this galaxy. However, for its metallicity \citep[$\sim 0.5 \times$ solar, according to][]{James2009A-VLT-VIMOS-stu} the C/O abundance ratio is comparable to that of other galaxies and \hii\ regions, hence not unusual.

Taken together we thus conclude that all of the six N-emitters show an elevated (supersolar) N/O abundance ratio, whereas the C/O abundance is normal in four of them, and only one of them (the Lynx arc) appears enhanced in C/O.
The observed and other properties of these objects are also summarized in Table~\ref{tab_scenarii}.
We will now discuss possible scenarios to explain to observed abundance pattern.

 \subsection{Possible chemical enrichment scenarios}
 
Galactic chemical evolution models are able to reproduce the observed average trends of the abundance ratios of CNO and H for ``normal'' galaxies \citep[see e.g.][]{Henry2000On-the-Cosmic-O,Chiappini2006,Berg2019The-Chemical-Ev,Johnson2023Empirical-const}, although the evolution of Nitrogen has notoriously been more complicated to explain since the observations show a behaviour like a primary element at low (subsolar) metallicity \cite[cf. discussion and references in][]{Prantzos2018MNRAS}.
To examine the conditions that may be more appropriate for low metallicity dwarf galaxies and \hii\ regions, which dominate the current samples of extra-galactic CNO measurements in galaxies (the samples shown here), various authors have studied the effects of variable or bursty star-formation histories, outflows, and different star-formation efficiencies. Again, such models are able to reproduce the {\em average} trends of C/O, N/O and C/N as a function of metallicity and they can also explain the observed scatter in the data, e.g.\ by the presence of burst phases \citep[see][for a recent study]{Berg2019The-Chemical-Ev}.

Since the observed abundance ratios of \source\ and possibly other N-emitters are, however, clearly more extreme than those of the bulk of galaxies studied so far, we need to examine the possible nucleosynthetic sources and the conditions capable of explaining them. To do so, we first consider two quantitative scenarios, the first involving enrichment from normal massive stars, and the second nucleosynthesis from super-massive stars. These scenarii were considered in recent studies \citep[e.g.][]{Charbonnel2023N-enhancement-i,Nagele2023Multiple-Channe,Watanabe2023EMPRESS.-XIII.-}.

\subsubsection{Enrichment from massive stars  -- ``WR-scenario''}

It is well-known that the stellar winds of massive stars can carry important amounts of newly-created elements such as He and  N (from H-burning, the latter resulting at the expense of C and O) or C and O (from He-burning); those elements appear at the stellar surfaces and are ejected by the winds 
during the so-called Wolf-Rayet (WR) phases,  with N enhanced in the WN
phase and C enhanced in the subsequent WC phase \citep{Maeder1983A&A}. The stellar wind yields depend strongly on the initial mass and metallicity of the stars,
and also on other properties such as stellar rotation and the efficiency of mixing in the stellar interiors, or their evolution in close binary systems \citep[e.g.][]{2012A&A...542A..29G,2015A&A...581A..15S,2022A&A...667A..58P}.

Using the recent stellar yields from \citet{Limongi2018} we have computed the cumulative stellar wind yields of a simple stellar population as a  function of time, for a \citet{Kroupa2002Sci} IMF, three different metallicities ([Fe/H]$=-2$, $-1$ and 0, respectively) and three different initial rotational velocities (V$_{\rm Rot}=0$, 150 and 300 km/s, respectively). The latter value of  V$_{\rm Rot}=$300 km/s  was adopted in \cite{Charbonnel2023N-enhancement-i} to discuss the observations of GN-z11.
Assuming that stars more massive than 20--25 \msun\ do not explode but collapse and become black holes \citep[see discussion in][]{Prantzos2018MNRAS}, the stellar ejecta have exclusively a wind composition for several million years. In the first couple of Myr, that composition is the original one of the stellar envelope, then it is dominated by  H-burning products and subsequently, by He-burning products.
To compare with observed abundance ratios \cite{Charbonnel2023N-enhancement-i} assumed a dilution of the wind ejecta with an equal amount of ISM. Here we assume no such mixing, thus maximizing the effect of the stellar winds on the composition. Physically, this may correspond to the situation where the winds of the previous O-star phase, operating for a few Myr, have opened a cavity in the ISM where the winds of the subsequent WR phase are expanding. Actually, there is mixture with pristine ISM material, since we include the winds released by all stars above 12 \msun \ and in the considered period of 8 Myr the stars less massive than 20 \msun \ do not reach the WR phase.

\begin{figure}
\centering
\includegraphics[width=0.48\textwidth]{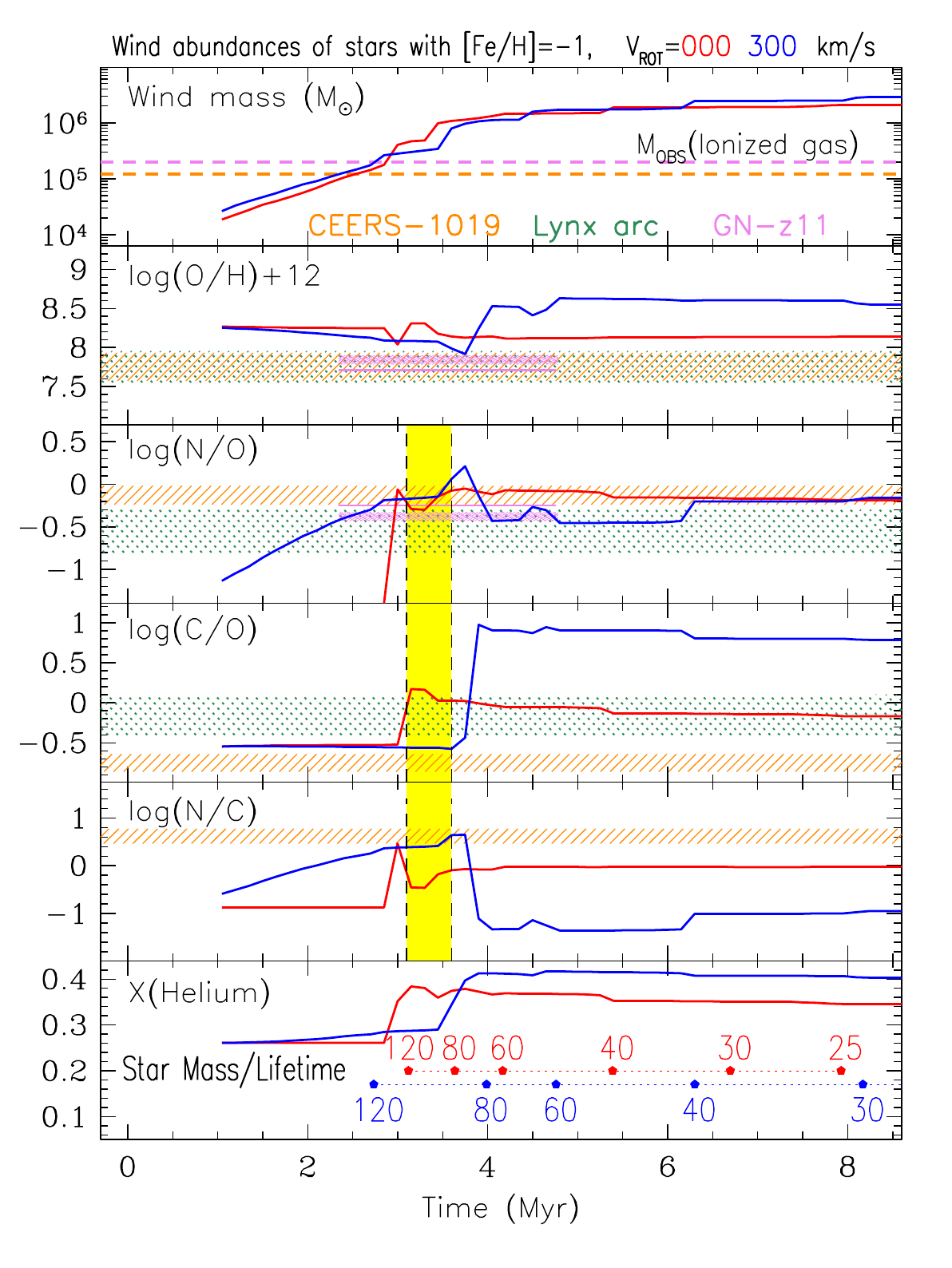}
\caption{Evolution of IMF-weighted time-integrated masses and abundances of the winds of a stellar population of total mass 10$^8$ \msun created at t=0, according to the models of \citet{Limongi2018} with metallicity [Fe/H]=-1 and initial rotational velocity V$_{\rm ROT}$=300 km/s (solid blue curves) or 0 km/s (solid red curves) in all panels;  practically no dilution with ambient ISM is assumed (99\% of ejecta and 1\% of ISM). Comparison is made to abundance data from CEERS-1019 (this work, orange shaded), Lynx arc (green shaded), GN-z11 \citep[][ violet shaded with age dermination]{Senchyna2023GN-z11-in-conte}. The two horizontal dashed lines in the top panel indicate the estimated mass of ionized gas observed in CEERS-1019 and GN-z11, respectively (see Sec. \ref{subsec:mass_gas_obs}).  The yellow shaded area indicates the short period ($\sim$0.5 Myr) where all three abundance ratios for CEERS-1019 are well reproduced by the rotating massive star winds. In the bottom panel, displaying the evolution of the He mass fraction, the corresponding lifetimes of the stars are indicated (filled circles, color-coded for V$_{\rm ROT}$= 0 and 300 km/s) for selected masses (associated numbers in \msun).
}
\label{fig_WRwinds}
\end{figure}

In Fig.\ \ref{fig_WRwinds} we display the evolution of various quantities of the "WR scenario" for stars of [Fe/H]$=-1$, a value reasonably close to the metallicity of the extragalactic systems studied here. Results are shown up to 8 Myr after the formation of a stellar population of  total mass $10^8$ \msun with a normal IMF \citep{Kroupa2002Sci}. During that period, stars below 25 \msun\ have not yet ended their lives (by assumption), so that only the wind ejecta populate the cavity crafted by the winds and the radiation of the stars. The mass of the wind ejecta increases steadily, from $\sim$10$^4$ \msun\ after the first Myr to $\sim$10$^6$ \msun\ at 4 Myr and more slowly after that. In Sec. \ref{subsec:mass_gas_obs} we discussed the amounts of ionized gas estimated in \source\ and GN-z11, which are compatible with the model results for this earliest period after the starburst (horizontal dashed lines in the top panel).

The evolution of the wind composition differs between the non-rotating and the rotating stars.
The former (solid red curves) have practically no mixing between their convective core and radiative envelope; in consequence, the signatures of H-burning (high N/O and N/C) appear abruptly in the wind, once the mass loss uncovers the former H-burning core. The latter (solid blue curves) undergo rotational mixing, bringing slowly the  H-burning products to the surface; as a  result, the N/O and N/C ratios increase slowly but steadily, up to the equilibrium value, which is similar to the case of non-rotating stars. The timescale for the appearance of high N abundance is $\sim 3$ Myr, in good agreement with the time window inferred by \cite{Senchyna2023GN-z11-in-conte} for GN-z11. About a Myr later, some amounts of He and He-burning products -- mainly C and insignificant  O amounts -- appear in the wind ejecta of the most massive rotating stars (from 120 to $\sim 70$ \msun) while the less massive ones never reach the WC phase; the combined effect is a strong increase of C/O, a strong decrease of N/C and a small variation of N/O. In contrast, none of the non-rotating stars reaches the WC phase at such low metallicity, and all the CNO ratios remain basically unchanged. After that, the situation is expected to change drastically, as the first SN from M$<$25 \msun \ stars explode and eject their core material in the ISM.

\begin{figure}[htb]
\centering
\includegraphics[width=0.49\textwidth]{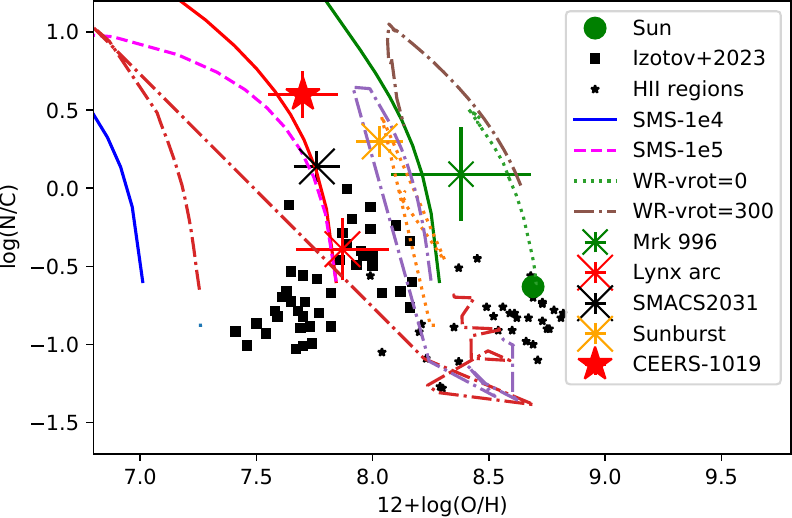}
\caption{Observed and predicted abundance ratios of N/C as a function of O/H for the six N-emitters and comparison samples from the literature.
Observed data are shown using the same symbols as in Fig.~\protect\ref{fig_abund}, model predictions with the linestyles of  Fig.~\protect\ref{fig_abund}.
}
\label{fig_models_nc}
\end{figure}

As shown in Fig.\ \ref{fig_WRwinds} in the early evolution of a stellar population, there is a period of several Myr during which the N/O ratio in the stellar winds reaches the high N/O ratios observed in \source\ and in the other N-emitters analyzed here.
However, rapidly after reaching the maximum  N/O value, the carbon abundance also increases (very strongly in rotating star or less so without rotation), implying C/O and N/C ratios that are incompatible with the observations of \source, SMACS2031, and the Sunburst cluster over most of the time (see also Fig.~\ref{fig_models_nc}). In the results displayed here, there is thus only a fairly short period of $\sim$ 0.5 Myr (yellow shaded area in Fig.\ \ref{fig_WRwinds}) where all three ratios N/O, N/C, and C/O are compatible with the observations of \source\ for the case of rotating stars. In view of the timescales involved (several Myr), the probability of such an occurrence is small but certainly non-negligible. We note that this occurs rather early in the evolution of the starburst, but well within the time window found by the analysis of \cite{Senchyna2023GN-z11-in-conte} for GN-z11 (violet horizontal segments in the 2nd and 3d panels). We also note that other stellar models than those used here could result in more extended periods of high N/O and N/C ratios. This could be the case, for instance, of stars rotating more rapidly than 300 km/s  \citep[e.g. the fast rotators at nearly break-up velocity of 800 km/s calculated by][]{Hirschi2007},  binary stars, or stars calculated with higher mass loss rates, etc. \citep[see e.g. ][for a recent review]{Eldridge2022ARA&A}. 
On the other hand, for the central region of Mrk 996 which shows both N and C enrichment, we find that all the abundance ratios are well reproduced by the models. Furthermore, in this galaxy the WR-scenario is directly supported by the presence of WR stars both of WN and WC types \citep{Telles2014A-Gemini/GMOS-s}. Similarly, N and C enrichment found in the Lynx arc could also be explained by the WR scenario, and earlier studies have argued for the presence of WR stars, from emission line modelling of this peculiar object \citep[see e.g.][]{Villar-Martin2004Nebular-and-ste}.

\begin{figure}[tb]
\centering
\includegraphics[width=0.49\textwidth]{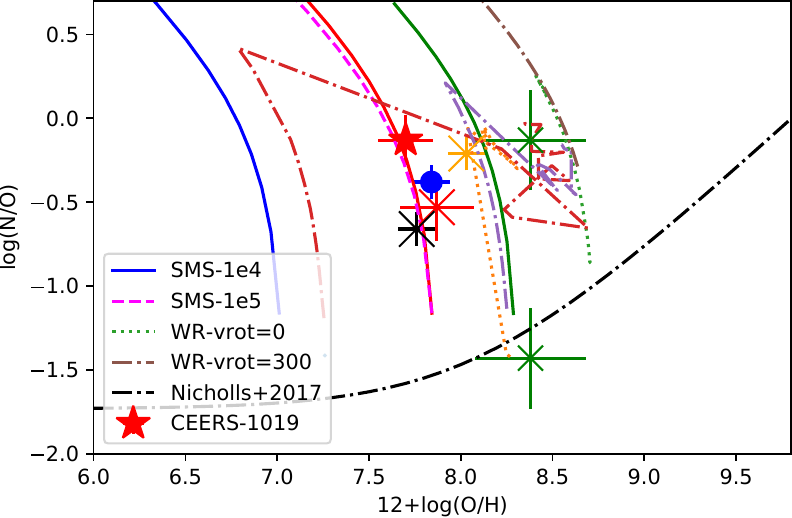}\vspace{0.5cm}
\includegraphics[width=0.49\textwidth]{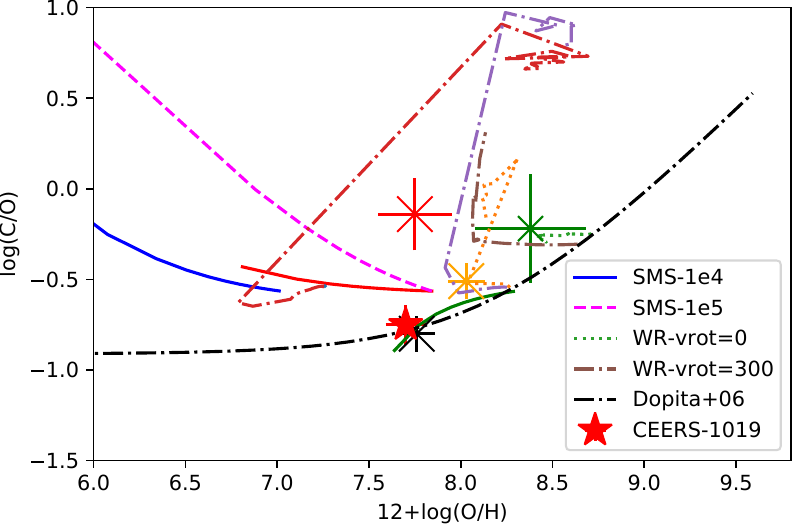}
\caption{Observed chemical abundances (N/O versus O/H in the top panel, C/O versus O/H in the bottom) of the six N-emitters (using the same symbols as in Fig.~\protect\ref{fig_abund}) and comparison with predictions for enrichment from massive stars (dotted and dash-dotted lines showing the ``WR scenario'' for non-rotating and rotating stars, respectively; see text) and supermassive stars (solid and dashed). Different colors indicate different metallicities. The predictions for the WR scenario are shown for a very low dilution (1\%) with ISM matter. The solid lines show the predicted composition using SMS models with 10$^4$ \msun\ at different metallicities (\oh $\approx $ 7.0,7.8, 8.3) from \cite{Charbonnel2023N-enhancement-i} and for varying amounts of dilution with an ISM of standard composition. The dashed lines show an SMS model with 10$^5$ \msun\ from \cite{Nagele2023Multiple-Channe}. }
\label{fig_models}
\end{figure}

Is there any direct evidence for WR stars in the N-emitters discussed here? In short, WR stars have been reported only in the low-$z$ galaxy Mrk 996, as mentioned earlier.
In the spectral range covered by the observations of \source, the strongest WR features could be \Heiiuv\ and \Civuv\ in the rest-UV and the so-called blue WR-bump centered around \Heiiopt. None of these features is detected in the current NIRSpec spectra and the same holds for GN-z11 \citep[see: ][]{Bunker2023JADES-NIRSpec-S,Maiolino2023_GNz11_AGN}. However, the JWST spectra of these very high-$z$ objects, and in particular for \source, are of insufficient spectral resolution and S/N to rule out, e.g., \Heiiuv\ emission with equivalents widths $\la 7-10$ \AA\ (depending on the adopted FWHM of the WR line), and therefore stellar populations comparable to those of Mrk 996, which has EW(1640)$ \approx 3-4$ \AA, cannot be ruled out from the present data. 
The rest-UV spectrum of SMACS2031 from  \cite{Patricio2016A-young-star-fo} also shows no clear feature of WR stars. \Heiiuv\ is present with an EW(1640)$=0.99 \pm 0.1$ \AA, but it is only marginally broader than nebular emission lines. 

The very high-S/N spectrum of the Sunburst cluster, discussed by \cite{Mestric2023Clues-on-the-pr}, also shows no signature of WR stars. Except for the nebular lines, the Sunburst spectrum resembles in fact strongly the spectrum of the well-known massive star cluster R136 in the LMC, which is known to be very young ($\sim 1.5$ Myr) and to host very massive stars with masses up to $\sim 200$ \msun\ \citep{Vanzella2020Ionizing-the-in,Mestric2023Clues-on-the-pr}. The Sunburst cluster also appears to be too young to host WR stars \citep{chisholm2019}. 
Finally, \cite{Villar-Martin2004Nebular-and-ste} have suggested the presence of WR in the Lynx arc, in particular to explain the hard observed ionizing spectrum, but no direct signatures are detected in the relatively low S/N spectra available for this object. 

In conclusion, except for Mrk 996 where the presence of important populations of WR stars (both of WN and WC types) is established, no direct evidence for WR stars is found in the other N-emitters studied here. However, this does not necessarily exclude the WR-scenario, since WR stars may be present below the detection threshold.

\subsubsection{Enrichment from super-massive stars ($M \protect\ga 1000$ \msun) -- SMS scenario}

An alternate scenario, already invoked by 
\cite{Charbonnel2023N-enhancement-i} to explain the high N-abundance in the compact galaxy GN-z11 at $z=10.6$ , is that of super-massive stars (SMS), which have previously been proposed to explain
the abundance anomalies of the multiple stellar populations seen in old Galactic and extra-galactic globular clusters (GC) and in extra-galactic massive star clusters with ages down to $\sim$ 1.7~Gyr
\citep{Gieles2018Concurrent-form}. 
In essence, this model proposes that gas accretion and collisions of proto-stars in the densest clusters lead to the runaway formation of one or several SMS, with masses $M \ga 10^3 \msun$ that increase with the cluster mass.
During some time before two-body relaxation heats the cluster, this mostly convective SMS undergoes accretion (from proto-stars in the cluster and infalling gas) and it ejects processed matter, whose composition reflects the conditions in its hot H-burning core.  Namely, the ejected material is strongly enriched in N, Na, and Al, and strongly depleted in O and C as a result of CNO, NeNa, and MgAl nuclear reactions at high temperature. As initially shown by \citet{Denissenkov2014}, the whole range of abundance anomalies (C-N, O-N, Na-O, Mg-Al anticorrelations) in GC stars is very well accounted for after dilution of the SMS ejecta with proto-GC gas. 

The constant supply of unprocessed material to the SMS ``freezes'' its evolution close to the zero-age main sequence, preventing strong He-enrichment of the SMS yields, in agreement with GC multiple band photometry \citep{2015MNRAS.446.1672M,2018MNRAS.481.5098M}. This also
solves the so-called ``mass budget" problem encountered by all the other scenarios that try to explain the presence and properties of multiple stellar populations in globular clusters \citep{2006A&A...458..135P,2011MNRAS.413.2297S,2012A&A...546L...5K,2015MNRAS.454.4197R,Krause2016_C5,Bastian2018Multiple-Stella}.
For example, \citet{Gieles2018Concurrent-form} find that a SMS forming into a dense cluster hosting $10^7$ proto-stars can reach and process respectively $\sim 5\%$ and $\sim 45\%$ of the cluster mass. This is significantly higher than the $\sim 2$\% of wind mass ejected in the massive star scenario (cf.\ Fig.~\ref{fig_WRwinds}).
In particular, the super-linear scaling predicted between the amount of material nuclearly processed by the SMS and the cluster mass explains the observed increase of the fraction of 
second population stars with GC mass \citep{2010A&A...516A..55C,2017MNRAS.464.3636M}. This picture is dubbed the ``conveyor-belt'' SMS model.
The high amount of processed matter also implies that any additional matter ejected by the SMS during its final phase (once the conveyer-belt stops) will have very little impact on the final abundance ratios.

In Figs.\ \ref{fig_models} and \ref{fig_models_nc} the solid lines show, for three different initial metallicities (0.34~Z$_{\odot}$, 0.12~Z$_{\odot}$, and 0.018~Z$_{\odot}$), the predicted chemical abundance ratios resulting from the mixture of ejecta of $10^4$~M$_{\odot}$ SMS in the conveyer-belt scenario with different amounts of ISM gas with a  normal, initial abundance (stellar models from \citealt{Charbonnel2023N-enhancement-i}). The composition of the SMS ejecta reflects the yields from H-burning via the CNO-cycle. It is very strongly enriched in Nitrogen, with N/O $>10$, i.e.\ nearly 100 times super-solar, and very strongly depleted in Oxygen and Carbon.
With an increasing fraction of matter from the SMS mixed into the ISM, the predicted N/O and N/C ratios increase strongly. The resulting mixture also shows a decreasing O/H abundance (metallicity) while C/O remains relatively constant.

The observed N/O ratio of GN-z11 and \source\ can be explained by mixing approximately equal amounts of SMS ejecta with ISM gas, as already shown by  \cite{Charbonnel2023N-enhancement-i}. The N/O abundance of all other N emitters considered here could also be explained with the SMS scenario. The same is also true for the C/O and N/C abundance ratios, except for the two objects which show a high C/O ratio, Mrk 996 and the Lynx arc. As already mentioned before, C/O in these galaxies reveals the presence of both H- and He-burning products, which, in the case of Mrk 996, is compatible with its observed WR star population. In short, the comparison of the observed N/O, C/O, and N/C ratios suggests that \source, SMACS2031, and the Sunburst cluster might be explained by the SMS conveyor-belt scenario, implying that they should contain one or several proto-GC, and Mrk 996 and the Lynx arc by the WR scenario. From the available data and the lack of accurate C/O measurements, the case of GN-z11 remains inconclusive. 

\cite{Nagele2023Multiple-Channe} have computed the composition of the material ejected through winds along the entire evolution of SMS with masses between $10^3$ and 10$^5$ \msun\  for 0.1~Z$_{\odot}$, neglecting the conveyor belt rejuvenation of the star discussed above (they assume that SMS form through gravitational collapse during the merger of gas-rich galaxies at high-$z$, see \citealt{2015ApJ...810...51M}). 
In addition, they estimate if and when the SMS become GR unstable as they evolve, as well as the modifications of the composition of the material that can be ejected at the end of the life of the stars in the case they eventually explode due to the CNO cycle and the rp (rapid proton capture) process (for details see \citealt{Nagele2023Multiple-Channe}). 
Their $10^3$ and $10^4$ \msun\ models -- not shown here -- predict strong N-enrichment on the main sequence, confirming the results of \citet{Denissenkov2014} and \cite{Charbonnel2023N-enhancement-i}. However, these two models do not become GR unstable and make it until carbon-oxygen burns. As a consequence, their winds reach super-solar C and O abundances because of the dredge-up of C and O from the core during central He-burning, and they are strongly enriched in He.
This implies that without undergoing the conveyor-belt episode that is required to solve the mass budget and the photometric constraints for the GC case, the total yields of such models cannot explain the GC abundance anomalies, nor can they explain the N/O and C/O ratios in CEERS-1019 
and in GN-z11 as discussed by \cite{Nagele2023Multiple-Channe}.

On the other hand, \cite{Nagele2023Multiple-Channe} find that their $5 \times 10^4$ and $10^5$ \msun\ models at 0.1~Z$_{\odot}$ become GR unstable close to or at the end of the main sequence, implying that their winds contain super-solar N and sub-solar C and O before the stars eventually collapse to a black hole or are disrupted by a thermonuclear explosion. 
The dashed lines in Figs.\ \ref{fig_models} and \ref{fig_models_nc} show the range of abundances expected when the ejecta of their $10^5$ \msun\ model is diluted to various degrees with ISM of the initial composition. In addition to the N-enrichment along the main sequence, this includes their estimate of the additional N that is produced during the expected CNO-powered explosion \citep{Nagele2023Multiple-Channe}.
This model accounts well for the observed abundance N/O ratios in \source, GN-z11, and SMACS2031.
It also produces enough enriched material to be able to pollute sufficient ionized gas, i.e.\ masses in the observed range (see Sect.~4.7), as shown by \cite{Nagele2023Multiple-Channe}.

From this, we conclude that SMS over a wide range of masses can simultaneously explain the GC abundance anomalies and the N/O and C/O ratios in CEERS-1019, GN-z11, and SMACS2031, if they eject large quantities of H-processed material early on the main sequence, as predicted by the conveyor-belt SMS scenario \citep{Gieles2018Concurrent-form}, or if the SMS sheds large amounts of processed material due to instabilities and an explosion during the CNO-cycle \citep[cf.][]{Nagele2023Multiple-Channe}.

In Sect.~\ref{s_protogc} we will further argue whether the N-emitters are proto-GCs, and discuss possible implications of the SMS scenario, including the possible formation of an intermediate-mass black hole (IMBH).

\subsubsection{Other scenarios to explain strong N emission}

\cite{Cameron2023Nitrogen-enhanc} have discussed different processes or sources that could explain the observed N-enhancement in GN-z11, including enrichment from AGB stars, pollution from Pop III star-formation, stellar encounters in dense star clusters, or tidal disruption of stars from encounters with black holes. The main conclusions of their qualitative discussion is that these explanations would need very fine-tuned conditions and that the origin of N-enrichment is currently largely unknown.

The predictions of classical chemical evolution models including AGB stars are shown e.g.\ in the studies of \cite{Johnson2023Empirical-const}. \cite{Watanabe2023EMPRESS.-XIII.-} also show predictions of such models in comparison with GN-z11. Indeed, as well known from earlier works, such models cannot produce high N/O abundance ratios at low metallicity
(as observed in the N-emitters discussed here), since these models include also the yields of massive stars and core-collapse supernovae, which produce large amounts of oxygen, and hence no extreme N/O ratios.
The pure WR-wind models of \cite{Watanabe2023EMPRESS.-XIII.-} are essentially the same as our massive star models (WR-scenario).
 
\cite{Maiolino2023_GNz11_AGN} have recently argued that GN-z11 shows signs of a type 1 AGN, with emission from very high density and a Broad Line Region (BLR). They further argue that the exceptionally high nitrogen abundance ``becomes much less problematic'' in the AGN scenario, for several reasons. First, they point out that several ``nitrogen-loud'' AGNs have been found, making GN-z11 less peculiar. And second, they mention that only small amounts of enriched gas are needed if the observed gas is at very high densities. Finally, they mention supernovae from supermassive stellar progenitors, rapidly recycled secondary nitrogen production, or bloated atmospheres of giant/supergiant stars as possible sources of the observed enrichment, without providing quantitative estimates.

Clearly, the spectra of \source\ and the other N-emitters discussed here are very different from nitrogen-loud AGN, as discussed in Sect.~\ref{s_agn}. Furthermore, except for GN-z11 for which \cite{Maiolino2023_GNz11_AGN} show indications of densities $n_H \ga 10^{10}$ \cmc\ typical of BLR, the densities inferred here are much lower, typically $n \sim 10^{4-5}$ \cmc, and all observed emission line properties are compatible with photoionization from star-formation (Sect.~\ref{s_agn}). The qualitative scenarios sketched by \cite{Maiolino2023_GNz11_AGN} for GN-z11 may therefore not be applicable to the other N-emitters discussed here. In any case, more quantitative studies on the detailed chemical abundances of nitrogen-loud AGN and their source of enrichment could be of interest to better understand the common points and differences with other N-emitters.

For the Sunburst cluster, \cite{Pascale2023Nitrogen-enrich} proposed a model where the super star cluster is surrounded by low- and high-density photoionized clouds and regions (channels) through which ionizing radiation can escape, and they argue that only the high-density clouds in the vicinity of the star cluster are N-enriched and confined by strong external pressure. They estimate that $\sim$ 500 \msun\ of nitrogen is needed -- an amount which can be produced by the star cluster with a mass $\mstar \sim$ few $\times 10^7$ \msun -- and suggest that it originates from young massive stars, ejected, e.g., in dense LBV winds or non-conservative binary mass transfer. SN ejecta are not favored, since the Sunburst is not enriched in C, and the inferred age ($\la 4$ Myr) is consistent with this explanation.

The model of \cite{Pascale2023Nitrogen-enrich} is essentially the same as our massive star scenario, although they do not use a specific model to predict the chemical yields of the cluster and its temporal evolution, and our massive star scenario does not include ejecta from mass transfer in binary systems. 
As already discussed above, such a scenario requires some specific ``tuning'', in particular the selection of a fairly specific age at which the composition of ejecta matches the observed abundances. For the Sunburst cluster, this seems very plausible; however, it is not clear if this could be generalized to \source\ and the other N-emitters.

\begin{table*}[ht]
\caption{Properties, proposed scenarii, and nature of the N-emitters }\label{tab_scenarii}
\begin{tabular}{lllllllllll}
\hline \hline
Object & redshift & N/O & C/O &  WR features  & enrichment  & proto-GC & BH formation & nature \\
\hline 
\source   & 8.678        & super-solar & normal       & ? & SMS  & yes & no  & SF-galaxy\\
GN-z11    & 10.6         & super-solar & uncertain    & ? & SMS or other & no & yes? & AGN? \\
SMACS2031 & 3.506        & super-solar & normal       & ? & SMS  & yes & no & SF-galaxy\\
Sunburst cluster & 2.368 & super-solar & normal       &   & SMS  & yes & no & SF-galaxy\\
Lynx arc  & 3.36         & super-solar & $\sim$ solar & ?     & WR? & no & no & WR galaxy?\\
Mrk 996   & 0.00544      & super-solar & $\sim$ solar & WC+WN & WR  & no & no & WR galaxy\\
\hline
\end{tabular}
\end{table*}

\subsection{Are \source\ and other N-emitters proto-GC in formation or related to the formation of intermediate-mass black holes ?}
\label{s_protogc}

The unusually high N/O abundances derived for GN-z11 and the Sunburst arc and similarities with the abundance pattern of stars in globular clusters have led several authors to suggest a link between these peculiar objects and GC formation \citep{Senchyna2023GN-z11-in-conte,Charbonnel2023N-enhancement-i,Nagele2023Multiple-Channe,Pascale2023Nitrogen-enrich}.
With the finding of a highly supersolar N/O ratio and normal C/O in \source\ and similar results for other objects from the literature (in total six N-emitters analyzed here), the question of the nature of the N-emitters must be rediscussed in light of new and additional evidence. We summarize basic observational evidence and our favourite scenarii/explanations in Table \ref{tab_scenarii}.

First, the observed abundance ratios of N/O and C/O, which are accurately measured for five objects, suggest that two of them (the low-$z$ galaxy Mrk 996 and the Lynx arc) are probably explained by pollution from WR stars, as discussed above. 
If correct, it implies that the cluster(s) dominating presumably these objects cannot be progenitors of GCs. This is due to the fact that massive star wind scenario suffers from the so-called mass budget problem of GCs \citep{2006A&A...458..135P,2007A&A...475..859D},  which basically means that the massive stars cannot produce sufficient amounts of enriched material to explain the observed population of ``polluted'' (second population) stars in GCs without this first population being much more massive than the second one, in contradiction with observations. 
In Mrk 996 WR features are detected, and the presence of WR stars is suspected in the Lynx arc. We therefore suggest that they are somewhat peculiar star-forming galaxies (WR galaxies), although we note that \cite{Binette2003High-z-nebulae} have also considered a hidden AGN to explain the emission line properties of the Lynx arc.

For \source, GN-z11, SMACS2031, and the Sunburst cluster, the N/O, C/O, and N/C ratios could be explained by the two scenarii discussed earlier, with the enriched matter originating from normal massive stars or from supermassive stars.
We favour the SMS scenario for several reasons. First, the WR scenario requires a very special and shorter timing than the SMS scenario. Second, these galaxies contain at least one sufficiently massive and compact region (the Sunburst cluster is of course a cluster) with extreme conditions (very high SFR and mass surface density), and unusually high ISM densities. Such rare conditions may be necessary for the formation of supermassive stars through runaway collisions and for the conveyer-belt model, as proposed by \cite{Gieles2018Concurrent-form}. This would also naturally explain why N-emitters are rare. We therefore propose that \source, SMACS2031, and the Sunburst cluster have been enriched by SMS and that they host (or are) proto-GCs in star-forming galaxies. Finally, the finding of such objects at look-back times between 11.2--13.3 Gyr is also compatible with them hosting proto-GCs.

The case of GN-z11 may be somewhat different as it may host an AGN, as suggested by \cite{Maiolino2023_GNz11_AGN}. If the high density of the ionized gas ($n_e \ga 10^{10}$ \cmc) inferred by these authors is confirmed, it would significantly reduce the amount of ionized gas that needs to be polluted, but it still leaves the source of chemical enrichment unexplained \citep[cf.][]{Maiolino2023_GNz11_AGN}. However, this does not exclude pollution from one or several SMS, which might even have seeded the ``small'' massive black hole (with $\log(M_{\rm BH}/\msun) \sim 6.2\pm0.3$) or contributed to its growth.
Indeed, the final fate of SMS is difficult to predict since in addition to metallicity and mass, other input parameters of the stellar models (mass loss, convection, overshooting, rotation, etc.) may impact the occurrence of the GR instability, its timing, and whether the collapse would trigger an explosion through the CNO-cycle \citep{1986ApJ...307..675F,2012ApJ...749...37M,2019A&A...632L...2H,2021A&A...650A.204H,Nagele2023Multiple-Channe}. In any case, the formation of IMBH with masses $\sim 10^4$ to 10$^5$ \msun\ from SMS seems possible at metallicities comparable to that of GN-z11, as shown e.g.\ by \cite{Nagele2023Evolution-and-e}. We therefore propose that N-emitters could also be an indication of black hole seed formation from SMS. And these objects could evolve to N-loud quasars, a rare sub-population of quasars showing strong N lines in the UV \citep{Jiang2008_N_QSOs}, and which have been suggested to be objects with high N/O and sub-solar metallicities in a rapid growth phase \citep{Araki2012Near-infrared-s,Matsuoka2017Chemical-enrich}.
We therefore mark GN-z11 as a possible AGN with BH-formation related to SMS in Table \ref{tab_scenarii}.
Finally, we also consider that the formation of an IMBH  
with mass $\ga 1000$ \msun~    
 from an SMS  is incompatible with the proto-GC scenario, as the presence of such a BH in old GCs seems to be ruled out observationally \citep[][and references therein]{Baumgardt2019}. This is also reflected in Table~\ref{tab_scenarii}.

Finally, we wish to remind the reader that \cite{Larson2023A-CEERS-Discove} suggested that \source\ also hosts a black hole, although our analysis does not show significant evidence for this and suggests that the object is dominated by star-formation (see Sect.~\ref{s_agn}). If \source\ harbours an AGN, the situation could be similar to that of GN-z11, just discussed and point to a possible link between SMS and black hole formation.
Also, we note that \cite{Binette2003High-z-nebulae} have considered a hidden AGN to explain the emission line properties of the Lynx arc, although the nature of this source remains unclear. 
To conclude, we also recall that none of the four other N-emitters discussed here show any AGN indication.
We are therefore probably left with three good candidates for SMS and proto-GCs, \source, SMACS2031, and the Sunburst cluster. 

\subsection{Future steps and improvements}

Clearly, better data and more N-emitters would be helpful to better understand the origin of the strong N emission lines, to further test the proposed enrichment scenarios and the possible presence of SMS, and thus to understand the nature of these rare objects. 

An important test for the massive star scenario would be to detect direct spectral signatures of WR stars. Deeper, very high S/N spectra, in the rest-optical domain would be ideal for this. 
In contrast to SMS, the massive star scenario also predicts important amounts of helium in the ejecta, which might be measurable from the analysis of nebular He and H emission lines in rest-optical spectra of sufficient quality.
In the SMS scenario, a strong enrichment of aluminum, originating from H-burning from the MgAl chain \citep{2007A&A...470..179P}, is predicted (Ramirez-Galeano, in prep.), as observed in GC stars \citep{2009A&A...505..139C,2017A&A...601A.112P,2019A&A...622A.191M}. In contrast, massive stars should produce less aluminum \citep{2007A&A...475..859D,2023A&A...673A.109G}.
Aluminum has spectral signatures in the rest-UV (Al~{\sc ii} $\lambda$1670, Al~{\sc iii} $\lambda\lambda$1855,1863), which are often seen in absorption in high-$z$ galaxy spectra \citep{shapley03,Le-Fevre2019The-VIMOS-Ultra}, and which are in emission in some AGNs \citep[see e.g.][]{Alexandroff2013_AGN}. These features might be an interesting test of the relation between N-emitters and proto-GCs, and to distinguish between the WR and SMS scenarii.

To examine if the strong N lines could be related to large density variations and found preferentially in pockets of high density, it will be of interest to obtain multiple density measurements probing the widest possible range of density, regions of different ionization, and possibly also spatial variations. Both high S/N and high-resolution spectra are needed for this, and measurements of fine-structure lines of oxygen and nitrogen with NOEMA 
could also provide insights into this question.

Future studies may reveal new N-emitters, improving the statistics and providing more test cases. If strongly enhanced N-emitters are found at significantly lower metallicities (say \oh $\la 7$) the SMS scenario might be favored, since WR stars should be less present at low O/H. Also, objects with even higher N/O abundances could exist, if the SMS scenario is correct.

\section{Conclusion}
\label{s_conclude}

In this work, we have presented the detailed analysis of \source\ at $z=8.678$ using deep spectroscopy and imaging with NIRSpec and NIRCam obtained from the \textit{JWST} CEERS program. Low- and medium-resolution NIRSpec spectra covering $1-5\mu$m reveal a wealth of rest-frame UV and optical nebular emission lines of various transitions and ionizing states from H, He, C, N, O, and  Ne. In particular, \source{ }shows remarkably intense Nitrogen emission of N~{\sc iii} and N~{\sc iv}, with N~{\sc iv}] $\lambda1486$ emerging as the strongest line within the rest-frame UV spectrum. These emission lines are very rarely seen in galaxy spectra, and \source\ -- which shows some resemblance with the peculiar object GN-z11 revealed recently by JWST \citep{Bunker2023JADES-NIRSpec-S} -- is thus the second ``N-emitter'' found at $z>8$. From the analysis of these data, we arrive at the following main results:

\begin{itemize}

    \item{Using the well-detected auroral [O~{\sc iii}] $\lambda4363$ line we determined the O/H abundance using the direct method, resulting in $\oh = 7.70 \pm 0.18$. We derived the electron temperature from both rest-frame UV and optical [O~{\sc iii}] lines, yielding consistent values of $T_{e} \approx 18000$~K. The density-sensitive lines of N~{\sc iv}]~$1483/1487 = 0.50\pm 0.22$ and [O~{\sc ii}]~$3727/3729=0.98\pm0.23$ suggest a relatively high electron density of $n_{e} \approx 10^{3-5}$ cm$^{-3}$. 
    These values are consistent with those reported by other studies for this object \citep{Tang2023JWST/NIRSpec-Sp, Nakajima2023JWST-Census-for, Larson2023A-CEERS-Discove}. }\vspace{2mm}

    \item{Metal abundances were derived for different ions of C, N, O, and Ne. Notably, we found an exceptionally high N/O abundance of log(N/O)$=-0.13\pm0.11$, approximately $5.6$ times higher than the solar ratio. Conversely, \source{ }exhibits relatively normal C/O and Ne/O ratios for its metallicity (O/H), with log(C/O)$=-0.75\pm 0.11$ and log(Ne/O)$=-0.63\pm0.07$, respectively. This translates to high N/O and N/C, and normal C/O ratios, typically found in globular cluster stars, and which reflect the abundance ratios from H-burning via the CNO-cycle at very high temperature \citep{2017A&A...608A..28P,2019A&ARv..27....8G}.
    }\vspace{2mm}

    \item{We have discussed possible chemical enrichment scenarios to explain these peculiar C, N, and O abundance ratios observed in \source. Enrichment from massive star winds through the WR phase can explain the observed ratios but requires a very short and specific time window (and the presence of WN stars only); it would also come with a very strong He enrichment. Furthermore, no signatures of WR stars are detected in \source, although their presence cannot be ruled out from the available data. Alternatively, models of super-massive stars ($>1000 M_{\odot}$) mixed with ISM with a normal composition can explain the abundance ratios of \source. In this scenario, the ejected processed material via SMS will exhibit H-burning products only, strong enriched in N and possibly some depletion in O and C, and a normal He content. 
    }\vspace{2mm}

    \item{We have investigated the possibility of an AGN in \source, a scenario recently suggested by \cite{Larson2023A-CEERS-Discove} due to the detection of a broad component in H$\beta$. Our own reduction of the NIRSpec spectrum shows a tentative, broad component in H$\beta$ ($\rm FWHM\simeq 1150$ km s$^{-1}$) but detected with a fairly low significance ($\simeq 2.2 \sigma$). Line ratios using rest-UV lines (N~{\sc v}, N~{\sc iv}], C~{\sc iv}, C~{\sc iii}], O~{\sc iii}], and He~{\sc ii}) suggest that the gas is primarily photoionized by star formation, and any contribution from an AGN would likely be residual. The non-detection of the high-ionization lines of N~{\sc v} $\lambda 1240$ and He~{\sc ii} $\lambda 1640$ further support this scenario. }\vspace{2mm}

    \item{\source{ }shows a complex morphology with three resolved clumps. By analyzing the light distribution of these substructures, we found very compact morphologies with characteristic half-light radii of $\simeq 100-150$~pc. Multi-wavelength SED fits for each individual clump predict stellar masses of log($M_{\star}/M_{\odot})\simeq 8.66-8.94$, resulting in very high stellar mass surface densities log($\Sigma_{M_{\star}}/(M_{\odot} \rm \, pc^{-2}) \simeq 3.55-4.14$. The star formation rate appears very intense in two clumps ($\rm SFR \simeq 80-150$ $M_{\odot}$ yr$^{-1}$), while the remaining clump displays a negligible level of ongoing star formation. }

\end{itemize}

\source{ }represents thus the second example of a rare population of strong N-emitting galaxies at $z>8$ with highly super-solar N/O abundances, very compact regions, and a high-density ISM.
To put this object into context and better understand these N-emitters, we have (re-)analyzed other known N-emitting star-forming galaxies from the literature. This includes three lensed objects, two galaxies (SMACS2031 and the Lynx arc), and one star-cluster (the Sunburst cluster) at $z \sim 2.3-3.5$, plus a nearby blue compact dwarf galaxy (Mrk 996), all of them without any clear indications of AGN activity. 
Similar to \source, these sources show peculiar abundance ratios with a supersolar N/O ratio along with very dense clustered mass and star formation (log($\Sigma_{M_{\star}}/(M_{\odot} \rm \, pc^{-2})) \gtrsim 3.5$) and high ISM densities ($n_e \sim 10^4-10^5$ \cmc). Two galaxies, Mrk 996 and the Lynx arc, show an enhanced C/O ratio compared to normal galaxies at the same metallicity (O/H), indicative of enrichment from WR stars. 

We have also presented quantitative predictions for the chemical enrichment in two different scenarios, including enrichment from winds of massive stars (called the WR-scenario) or from ejecta of supermassive stars (SMS) with masses $10^3-10^5$ \msun, which have been invoked to explain the abundance anomalies observed in present-day globular clusters \citep{Denissenkov2014,Gieles2018Concurrent-form}. The WR scenario explains well the two galaxies with enhanced C/O and is supported by direct evidence of WN and WC stars in Mrk 996. As already found by \cite{Charbonnel2023N-enhancement-i} for GN-z11, we found that the SMS scenario reproduced well the observed abundance ratios in \source, SMACS2031, and the Sunburst cluster. 
These observations probably provide the best indirect evidence so far for the possible existence of SMS in galaxies.

Finally, considering the preferred enrichment scenarii and other physical properties, we have also examined which of the N-emitters could host proto-GCs and what their nature is. From our analysis we concluded that \source, SMACS2031, and the Sunburst cluster host most likely proto-GCs. We also suggested that the peculiar abundances of GN-z11 could be due to SMS, even if this object was confirmed to host an AGN, as proposed by \citep[see][]{Maiolino2023_GNz11_AGN}. This could also point to the formation of intermediate-mass black holes from SMS and suggest a link between the N-emitters and N-loud quasars.

In short, the newly discovered N-emitter \source\ and other N-emitters show tantalizing similarities with stars in GCs and the conditions expected during the formation of GCs. They may also offer a unique window into the formation of SMS, their role during the formation of GCs, and also their possible importance as seeds for the formation of massive black holes. More detailed studies and further discoveries of these rare objects will shed further light on these exciting topics and questions.

\begin{acknowledgements}

We would like to thank the referee for a thoughtful report that improved the manuscript. 
We thank Lise Christensen and Johan Richard for sharing spectra from their VLT observations of SMACS2031.
We also thank Mark Gieles, Eros Vanzella, Laura Ramirez Galeano, Anastasios Fragos, Holger Baumgardt, Montse Villar-Martin and other colleagues for stimulating discussions.
CC acknowledges support from the Swiss National Science Foundation (SNF; Project 200020-192039). M.M. acknowledges the support of the Swedish Research Council, Vetenskapsr\r{a}det (internationell postdok grant 2019-00502).
Y.I. acknowledges support from the National Academy of Sciences of Ukraine (Project No. 0123U102248) and from the Simons Foundation.

\end{acknowledgements}
\bibliographystyle{aa}
\bibliography{references}

\end{document}